\documentclass[aps,showpacs,floatfix,superscriptaddress,amsmath,amssymb,twocolumn]{revtex4-1}
\usepackage{newtxtext,newtxmath}   
\usepackage{color}
\usepackage[pdftex]{graphicx}
\usepackage[pdftex,colorlinks]{hyperref}
\usepackage{amsmath}
\usepackage{float}
\usepackage{bbold}
\usepackage{todonotes}
\usepackage[caption=false]{subfig}

\begin{document}

\title{ 
{Open loop linear control of quadratic Hamiltonians with applications} 
}

\author{ Mattias T. Johnsson }	
\affiliation{Department of Physics and Astronomy, Macquarie University, North Ryde, NSW 2109, Australia}

\author{Daniel Burgarth}	
\affiliation{Physics Department, Friedrich-Alexander Universit{\"{a}}t of Erlangen-Nuremberg, Staudtstr. 7, 91058 Erlangen, Germany}

\begin{abstract}
The quantum harmonic oscillator is one of the most fundamental objects in physics. We consider the case where it is extended to an arbitrary number modes and includes all possible terms that are bilinear in the annihilation and creation operators, and assume we also have an arbitrary time-dependent drive term that is linear in those operators. Such a Hamiltonian is very general, covering a broad range of systems including quantum optics, superconducting circuit QED, quantum error correcting codes, Bose-Einstein condensates, atomic wave packet transport beyond the adiabatic limit and many others. We examine this situation from the point of view of quantum control, making use of optimal control theory to determine what can be accomplished, both when the controls are arbitrary and when they must minimize some cost function. In particular we develop a class of analytical pulses. We then apply our theory to a number of specific topical physical systems to illustrate its use and provide explicit control functions, including the case of the continuously driven conditional displacement gate.
\end{abstract}

\maketitle

\section{Introduction}
Broadly speaking control theory is concerned with the inverse problem of engineering differential equations to achieve target evolutions. In quantum control, the differential equations are usually the Schr{\"{o}}dinger equation for the noiseless or master equations for the noisy case, although other types of evolutions can also be considered. One can distinguish open loop control \cite{dalessandro2008,elliot2009,agrachev2004}, where no measurement information is used for controls, and closed loop control \cite{nurdin2018,wiseman2009}, where measurements results are fed back to determine future controls. The first case leads to a beautiful geometric theory, while the latter has an intricate stochastic aspect. Just as in classical control theory, analytical solutions are scarce, and one often uses numerical optimisation. For closed loop control, the special case of harmonic oscillator systems with linear control has received much attention due to its physical relevance and due to its analytical tractability. The purpose of the current study is to develop the general theory of open loop control of quadratic Hamiltonians and to apply this to physically relevant setups. The theory draws mostly from classical results \cite{zabczyk2020} and allows us to generalise the pioneering work \cite{mirrahimi2004} and to derive a deeper understanding of specific analytical control methods \cite{calarco2009}.

The structure of this paper is as follows. In Section~\ref{secQHOandNotation} we set up the problem, define notation, and extend the multimode quadratic Hamiltonian to include arbitrary time-dependent linear control terms. This system is mapped to a symplectic form, and shown to be equivalent to a classical linear control problem provided we consider only expectation values of the mode operators. Section~\ref{secQuantumAndClassicalControlTheory} provides a brief overview of classical linear control and introduces the concept of controllability and the Kalman matrix. We provide a derivation of the Kalman controllability criterion and show how to construct explicit control functions that steer the system from a given initial state to a given final state. This material is not new, but it is difficult to find a clear, concise and self-contained exposition in the standard literature. We also extend these results to complex state state spaces and control functions, which are required for our quantum formulation, but do not appear to be considered in standard texts.

Given this background, in Section~\ref{secControllingTheQBH} we investigate the limits of our control on the system, demonstrating that we can achieve multimode displacement operations, but not, for example, squeezing. Due to the results in Section~\ref{secQuantumAndClassicalControlTheory} we are able to find the explicit time-dependence required for the control functions to perform any such displacements. We also provide physical insight into our results by showing how the controllability of such a system is related to the normal modes of the quadratic Hamiltonian, as well as explicitly considering an example examining the controllability of linear bosonic chain with nearest neighbour couplings that include squeezing terms.

Section~\ref{secOptimalControl} extends our results to the case of optimal control, where we not only wish to find control functions that move our system from one configuration to another, but also minimize some penalty cost while doing so, such as time or energy. We illustrate our results by considering a two-mode system with squeezing terms, and find a explicit control function that steers the system while prioritizing either keeping the strength of the control low, or keeping the system close to the origin.

Finally, in Section~\ref{secApplications} we show how our results can be applied to four very different physical systems that are currently of interest in the areas of quantum information and quantum control. Specifically, we consider the following. 1) We begin by examining the situation where one uses a harmonic potential trap to move an atomic wave packet from one location to another, at a speed far beyond the adiabatic limit, that nonetheless preserves the orginal wave function exactly. Our results show that we can do as well as previously suggested control methods, but also allows additional constraints such as bounding the amplitude of the control function. Furthermore, our method allows the easy generation of an arbitrary number of explicit control functions with different properties, which can be selected to be experimentally easier to achieve. 2) The echoed conditional displacement gate. We show how to find a control pulse that conditionally displaces a coherent bosonic cavity field by $\pm \beta$ depending on the state of a qubit. Our example does not include optimal control, but it could easily be added if one requires additional constraints on the controls or state of the system during the gate operation in order to reduce decoherence, for example. 3) Optomechanics. We show how our formalism allows finding explicit control functions that control the dynamics of a cavity-coupled micromechanical oscillator by adjusting the number of photons in the cavity as a function of time, which in turn is equivalent to the altering the easily controlled parameter of laser power. 4) Control of cavity modes in the context of circuit quantum electrodynamics (QED). QED allows, for example, the implementation of superconducting qubits. Such qubits, however, require the presence of a nonlinearity to change the evenly spaced harmonic energy levels into an anharmonic spectrum where the lowest two levels can be treated as an independent system. This nonlinearity would seem to fall outside our linear scheme, but we show how one can use optimal control techniques to minimize the effect of such a nonlinearity, and thus still find analytic control functions that successfully manipulate the system.

\section{The quadratic bosonic Hamiltonian and notation}
\label{secQHOandNotation}

The most general Hamiltonian that is at bilinear in terms of bosonic annihilation and creation operators is given by 
\begin{equation}
\hat{H} = \sum_{i,j} \left( G_{ij} \hat{a}^{\dagger}_i \hat{a}_j + \frac{1}{2} B_{ij} \hat{a}^{\dagger}_i \hat{a}^{\dagger}_j +  \frac{1}{2} B^{\ast}_{ij} \hat{a}_j \hat{a}_i \right) 
\label{eqFullQuadraticControlHamiltonian}
\end{equation}
where $G$ and $B$ are square matrices with elements indexed by $i,j$. As $\hat{H}$ is Hermitian, we have the relations $G=G^{\dagger}, B=B^{T}$.

This Hamiltonian describes a system with an arbitrary number of coupled harmonic oscillator modes given by operators $\hat{a}_i$ satisfying the bosonic commutation relations $[\hat{a}_i, \hat{a}^{\dagger}_j] = \delta_{ij}$. Such a quadratic bosonic Hamiltonian is ubiquitous in quantum physics, both as a exact Hamiltonian and in effective theories, appearing in ion trapped quantum computing, optomechanics, superconducting circuits, quantum field theories and many other areas, and has been thoroughly studied.


Our interest is in extending the Hamiltonian (\ref{eqFullQuadraticControlHamiltonian}) to include arbitrary time-dependent affine coupling terms, giving 
\begin{eqnarray}
\hat{H} &=& \sum_{i,j} \left( G_{ij} \hat{a}^{\dagger}_i \hat{a}_j + \frac{1}{2} B_{ij} \hat{a}^{\dagger}_i \hat{a}^{\dagger}_j +  \frac{1}{2} B^{\ast}_{ij} \hat{a}_j \hat{a}_i \right) \nonumber \\
&& \,\,\,\, + \sum_i \left( c^{\ast}_i(t) \hat{a}_i + c_i(t) \hat{a}_i^{\dagger} \right) 
\label{eqFullQuadraticControlHamiltonianWithLinearTerm}
\end{eqnarray}
where the $c_i(t)$ are per-mode time-dependent functions. While these coupling terms are completely general, we will later assume they correspond to controls we can apply to the system in order to control its state. Specific physical scenarios will be discussed in Section \ref{secApplications}, but we do note that this setup is to be contrasted with the bilinear control of the quantum harmonic oscillator considered in \cite{Genoni}.


We now introduce vectors of the bosonic modes, as well a vector describing the linear couplings. Specifically we define
\begin{eqnarray}
\hat{{\boldsymbol{\alpha}}} &=&  
\begin{bmatrix}
\hat{a}_1 &
\hat{a}_2 &
\cdots &
\hat{a}^{\dagger}_1 &
\hat{a}^{\dagger}_2 &
\cdots
\end{bmatrix} ^T, \,\,\,\, \\
{\mathbf{c}}(t) &=&  
\begin{bmatrix}
c_1(t) &
c_2(t) &
\cdots &
c_1^{\ast}(t) &
c_2^{\ast}(t) &
\cdots
\end{bmatrix} ^T,
\label{eqVecAdefinition} 
\end{eqnarray}

\begin{eqnarray}
{\hat{\boldsymbol{\alpha}}^{\dagger}} &=&  
\begin{bmatrix}
\hat{a}^{\dagger}_1 &
\hat{a}^{\dagger}_2 &
\cdots &
\hat{a}_1 &
\hat{a}_2 &
\cdots
\end{bmatrix}, \\
{\mathbf{c}}^{\dagger}(t) &=&  
\begin{bmatrix}
c_1^{\ast}(t) &
c_2^{\ast}(t) &
\cdots &
c_1(t) &
c_2(t) &
\cdots
\end{bmatrix},
\end{eqnarray}
such that $\hat{{\boldsymbol{\alpha}}}$ and ${\mathbf{c}}(t)$ are column vectors of operators and $\hat{{\boldsymbol{\alpha}}}^{\dagger}$ and ${\mathbf{c}}^{\dagger}$ are row vectors of operators. The Hamiltonian (\ref{eqFullQuadraticControlHamiltonian}) can now be written as \cite{blaizot1986,derezinski2017}
\begin{equation}
\hat{H} = \frac{1}{2} \hat{{\boldsymbol{\alpha}}}^{\dagger} M \hat{{\boldsymbol{\alpha}}} -\frac{1}{2} {\text{Tr}}(G) + {\mathbf{c}}^{\dagger}(t) \hat{{\boldsymbol{\alpha}}}
\label{eqSymplecticQuadraticH}
\end{equation}
where the matrix $M$ is Hermitian and is given in block form as
\begin{equation}
M =   \begin{bmatrix}
G & B \\
B^{\ast} & G^{\ast}
\label{eqMdefinition}
\end{bmatrix}.
\end{equation}

We now derive the Heisenberg equations of motion for a specific bosonic mode $\hat{a}_i$ in the Heisenberg picture. This is standard, but usually not done in the case of affine terms. In what follows we will not indicate the time dependence of the Heisenberg picture operators $\hat{\boldsymbol{\alpha}}$ explicitly unless we wish to be especially clear. We find
\begin{eqnarray}
i \hbar \frac{d}{dt} \, \hat{a}_i &=& [\hat{a}_i, \hat{H}(t)]  \label{eqHeisenbergEOMrelation}\\
&=& \sum_{kl} \left( G_{kl} [\hat{a}_i , \hat{a}_k^{\dagger} \hat{a}_l ] + \frac{1}{2} B_{kl} [\hat{a}_i , \hat{a}_k^{\dagger} \hat{a}^{\dagger}_l ] \right) \nonumber \\
&& + \sum_k [\hat{a}_i,  c_k(t) \hat{a}_k^{\dagger}].
\end{eqnarray}
Using $[\hat{a}_i, \hat{a}_k^{\dagger}] = \delta_{ik}$, $[\hat{a}_i, \hat{a}_k^{\dagger} \hat{a}_l] = \delta_{ik} \hat{a}_l$, and $[\hat{a}_i, \hat{a}_k^{\dagger} \hat{a}^{\dagger}_l] = \delta_{il} \hat{a}_k^{\dagger} + \delta_{ik} \hat{a}_l^{\dagger}$ we obtain
\begin{eqnarray}
i \hbar \frac{d}{dt} \, \hat{a}_i &=& \sum_{kl} \left( G_{kl} \delta_{ik} \hat{a}_l + \frac{1}{2} B_{kl} (\delta_{il} \hat{a}_k^{\dagger} + \delta_{ik} \hat{a}_l^{\dagger}) \right) +  \sum_k \delta_{ik} c_k(t) \nonumber \\
%
 %
  &=& \sum_{l} G_{il} \hat{a}_l + \sum_{l} B_{il} \hat{a}_l^{\dagger} +   c_i(t) \label{eqHEOMforai}
\end{eqnarray}
where we used the fact that $B_{ij} = B_{jk}$. As the Hermitian transpose of this equation is given by
\begin{equation}
i \hbar \frac{d}{dt} \, \hat{a}^{\dagger}_i = -\sum_{l} G_{il}^{\ast} \hat{a}^{\dagger}_l - \sum_{l} B^{\ast}_{il} \hat{a}_l - c_i^{\ast}(t).
\label{eqHEOMforaidagger}
\end{equation}
we can write Eqs.~(\ref{eqHEOMforai}) and (\ref{eqHEOMforaidagger}) combined in matrix form as
\begin{equation}
i \hbar \frac{d}{dt} \, \hat{{\boldsymbol{\alpha}}} = 
\begin{bmatrix}  
G & B \\
-B^{\ast} &-G^{\ast}
\end{bmatrix} 
\hat{{\boldsymbol{\alpha}}} +
\begin{bmatrix}  
c_i  \\
-c_i^{\ast}
\end{bmatrix}
\end{equation}
and, recalling our definition of the matrix $M$, we see this can be written as
\begin{equation}
\frac{d}{dt} \, \hat{{\boldsymbol{\alpha}}}(t) = -i \eta M \hat{{\boldsymbol{\alpha}}}(t) - i \eta {\mathbf{c}}(t)
\label{eqSymplecticDETimeDependent}
\end{equation}
where we have set $\hbar = 1$ and $\eta$ is the matrix \begin{equation}
\eta = [\hat{{\boldsymbol{\alpha}}} , \hat{{\boldsymbol{\alpha}}}^{\dagger} ] = \begin{bmatrix}
\mathbb{1} & 0 \\
0 & -\mathbb{1}
\end{bmatrix}.
\end{equation}

It is of course possible to embed the affine structure by an extended matrix. To this end, we augment our vector $\hat{{\boldsymbol{\alpha}}}$ with a single additional entry, in which case (\ref{eqSymplecticDETimeDependent}) can be written as
\begin{equation}
\frac{d}{dt} \, 
\begin{bmatrix}
\hat{{\boldsymbol{\alpha}}} (t)\\
1 
\end{bmatrix}
= 
 \begin{bmatrix}
-i \eta M & -i \eta  {\boldsymbol{c}} (t) \\
{\boldsymbol{0}} & 0
\end{bmatrix}
\begin{bmatrix}
\hat{{\boldsymbol{\alpha}}} \\
1
\end{bmatrix}.
\end{equation}

Although we have formulated the problem in terms of the mode operators $\hat{a}_i, \hat{a}^{\dagger}_i$, it is sometimes convenient to work with the position and momentum operators $\hat{x}_i, \hat{p}_i$ instead. These are connected through the standard transformations
\begin{eqnarray}
\hat{x}_i &=& \frac{1}{\sqrt{2}} \left(  \hat{a}_i + \hat{a}^{\dagger}_i \right), \\ 
\hat{p}_i &=& \frac{i}{\sqrt{2}} \left( \hat{a}^{\dagger}_i - \hat{a}_i \right).
\end{eqnarray}
We choose the vector of operators that fully characterises the system in this basis as
\begin{equation}
\hat{\boldsymbol{\beta}} =  
\begin{bmatrix}
\hat{x}_1 &
\hat{x}_2 &
\hat{x}_3 &
\cdots &
\hat{p}_1 &
\hat{p}_2 &
\hat{p}_3 &
\cdots
\end{bmatrix}^T
\end{equation}
and define the unitary
\begin{equation}
U_{\beta\alpha} = \frac{1}{\sqrt{2}} \begin{bmatrix}
{\mathbb{1}}  & {\mathbb{1}} \\
-i{\mathbb{1}}  & i{\mathbb{1}} 
\end{bmatrix}.
\end{equation}
The transformation between the $\hat{a}_i, \hat{a}^{\dagger}_i$ and $\hat{x}_i, \hat{p}_i$ basis is now given by
\begin{equation}
\hat{\boldsymbol{\beta}} = U_{\beta\alpha} \hat{\boldsymbol{\alpha}}, 
\end{equation}
and, in this new basis, the equations of motion are given by 
\begin{equation}
i \hbar \frac{d}{dt} \, \hat{{\boldsymbol{\beta}}} = U_{\beta\alpha} \, \eta M \, U_{\beta\alpha}^{\dagger}\hat{{\boldsymbol{\beta}}} +  U_{\beta\alpha} \eta {\mathbf{c}}(t)
\label{eqSymplecticDEUtransform}
\end{equation}


A formal solution to Eq.~(\ref{eqSymplecticDETimeDependent}) is straightforward, taking the form of the standard Cauchy solution. Specifically, the solution is given by
\begin{equation}
\hat{{\boldsymbol{\alpha}}}(t) = e^{-i \eta M t} \hat{{\boldsymbol{\alpha}}}(0) - i \int_0^{t} e^{-i \eta M (t-s)}  \eta {\mathbf{c}}(s) \, ds,
\label{eqFormalCauchySolution}
\end{equation}
as can be seen by taking the time derivative of (\ref{eqFormalCauchySolution}) via the Leibnitz integral rule and comparing with the right hand side of (\ref{eqSymplecticDETimeDependent}).
%
%
%

There are two simple limiting cases to Eq.~(\ref{eqFormalCauchySolution}) worth noting. The first is when ${\mathbf{c}}(t) = {\mathbf{0}}$, so the Hamiltonian has no linear coupling and is time-independent. In this case the solution is simply
\begin{equation}
\hat{{\boldsymbol{\alpha}}}(t) = e^{- i \eta M t} \hat{{\boldsymbol{\alpha}}}(0),
\label{eqFormalCauchySolutionf0}
\end{equation}
and provides an explicit analytic solution to the problem of the quadratic bosonic Hamiltonian in terms of the symplectic generator $\eta M$.

The second case is where the Hamiltonian has a constant linear coupling that is time-independent, so that ${\mathbf{c}}(t) = {\mathbf{c}} = {\text{constant}}$. In this case we can write the analytic solution to (\ref{eqFormalCauchySolution}) as
\begin{equation}
\hat{{\boldsymbol{\alpha}}}(t) = e^{-i \eta M t} \hat{{\boldsymbol{\alpha}}}(0) - (\eta M)^{-1} (\mathbb{1} - e^{-i \eta Mt}) \eta {\mathbf{c}}.
\label{eqFormalCauchySolutionf1}
\end{equation}

We now consider the case where we are interested only in the evolution of the {\emph{expectation values}} of the mode operators for the system. We proceed by considering the expectation values of Eq.~(\ref{eqSymplecticDETimeDependent}), obtaining
\begin{equation}
\frac{d}{dt} \, \langle \hat{{\boldsymbol{\alpha}}} \rangle = -i \eta M \langle \hat{{\boldsymbol{\alpha}}} \rangle - i  \eta {\mathbf{c}}(t)
\label{eqSymplecticDETimeDependentExpectation}
\end{equation}
and
\begin{equation}
\langle \hat{{\boldsymbol{\alpha}}}(t) \rangle  = e^{-i \eta M t} \langle \hat{{\boldsymbol{\alpha}}}(0) \rangle  -i \int_0^{t} e^{-i \eta M(t-s)} \eta  {\mathbf{c}}(s) \, ds.
\label{eqAlphaExpectationSolution}
\end{equation}
Crucially, all operators have now been removed, and we have equations of motion for the $\langle \hat{a}_i \rangle$ that consist only of vectors and matrices of complex numbers.

What is even more important, and a central point of this paper, is that Eq.~(\ref{eqSymplecticDETimeDependentExpectation}) has now taken the form of a classical \emph{linear} control problem, which is well-studied and understood. Understanding the reachability of the system allows us to know to what extent we can steer states and evolution operators of the system by specifying the control functions ${\mathbf{c}}(t)$. In order to do this, we will need the basics of  classical linear control theory, which we will now briefly summarize.

\section{Linear control theory}
\label{secQuantumAndClassicalControlTheory}

The addition of the arbitrary time-dependent linear term to the standard bosonic quadratic Hamiltonian (\ref{eqFullQuadraticControlHamiltonian}) means that we can now examine it from the point of view of quantum control.

The starting point for linear control theory \cite{zabczyk2020} is the linear system described by
\begin{equation}
\frac{d}{dt} {\mathbf{x}}(t) = A {\mathbf{x}}(t)  + C {\mathbf{u}}(t)
\label{eqLinearControlDefinition}
\end{equation}
$A$ is a complex $n\times n$ matrix and $C$ is a complex $n\times m$ matrix, so that $A: \mathbb{C}^n \rightarrow \mathbb{C}^n$, \, $C: \mathbb{C}^m \rightarrow \mathbb{C}^n$.  ${\mathbf{x}(t)} \in \mathbb{C}^n$ and ${\mathbf{u}}(t) \in \mathbb{R}^m$ are vectors, where the ${\mathbf{x}}(t)$ denote the $n$ system configuration variables at time $t$, and the ${\mathbf{u}}(t)$ denote the $m$ continuous real scalar control functions that we are able manipulate.

In many cases one specializes to a single control function ${\mathbf{u}}(t) = u(t)$, so that the matrix $C$ a becomes a single control vector ${\mathbf{c}}$, but intially we consider things completely generally.



The complex vector ${\mathbf{x}}(t)$ represents the state of the system at any given time. A system that evolves according to Eq.~(\ref{eqLinearControlDefinition}) is called controllable at time $T>0$ if we can move any initial state ${\mathbf{x}}(0)$ to any other target state ${\mathbf{x}}(T)$ at time $T$ by choice of suitable control functions ${\mathbf{u}}(t)$.

To determine whether a system is controllable, one considers the Kalman matrix $K$, constructed column-by-column via
\begin{equation}
K =: \left[ C \,\,\,\, AC \,\,\,\,  A^2C \,\,\,\, \dots \,\,\,\, A^{n-1}C   \right].
\label{eqKalmanMatrixDefinition}
\end{equation}
The Kalman rank criterion states that a system is controllable at a time $T>0$ if and only if $K$ has rank $n$. This also means that if a system is controllable, it is controllable for any time $T>0$, no matter how short the time, although this relies on the unbounded nature of the control functions ${\mathbf{u}}(t)$. If we place bounds or contraints on the magnitude of the control functions this is no longer necessarily true.

Suppose the system described by Eq.~(\ref{eqLinearControlDefinition}) is controllable. In this case not only do we know it is possible to steer any initial state to any final state, but it is also possible to analytically construct a set of control pulses ${\mathbf{u}}(t)$ that perform that evolution. In fact, there is freedom in choosing such a pulse, so it is possible to construct a control solution that has properties that are specifically useful to the system of interest. Although the Kalman controllability criterion is well-known, its extension to the complex domain and the creation of specific solutions is less so, and for that reason we present a proof in Appendix \ref{secAppendixKalmanProof}, and merely state the result here.

The explicit control functions that take ${\mathbf{x}}(0)$ to ${\mathbf{g}}$ at time $T$ are given by
\begin{equation}
{\mathbf{u}}(t) = \sum_{l=1}^n \bar{K}_l  \frac{d^{l-1}}{dt^{l-1}} {\mathbf{r}}(t)
\label{eqAnalyticSolutionForft}
\end{equation}
where the $\bar{K}_l$ is defined by noting that if $K$ has rank $n$, then there exists an $mn\times n$ matrix $\bar{K}$ with the property $K\bar{K} = \mathbb{1}_n$, where $\mathbb{1}_n$ denotes the $n\times n$ identity matrix. $K_l$ is given by the $n\times m$ sublock of $\bar{K}$, beginning at row $n(l-1)$. In the case where we have only a single control function, $K$ is square, and the $\bar{K}_l$ correspond to the rows of $K^{-1}$. ${\mathbf{r}}(t)$ is given by
\begin{equation}
 {\mathbf{r}}(t) = \mu(t) e^{A(t-T)} \left( {\mathbf{g}} -e^{AT}  {\mathbf{x}}(0) \right)
 \label{eqroftdefinition}
\end{equation}
and $\mu(t)$ is an auxiliary `bump' function that is chosen to satisfy the following criteria:
\begin{itemize}
\item The first $n-1$ derivatives of $\mu(t)$ are continuous
\item $\frac{d^l}{dt^{l}} \mu(t) = 0$ at $t=0$ and $t=T$ for $l=0,1,\ldots , n-1$
\item $\int_0^T \mu(t) \,dt = 1$.
\end{itemize}
One simple example of such a function is given by
\begin{equation}
\mu(t) = N t^n (T-t)^n
\label{eqBumpFunctionDefinition}
\end{equation}
where $N$ is a normalization constant, but any function that satisfies these criteria will work.

It is this freedom to choose ${\mathbf{u}}(t)$ that allows us to choose controls that may more closely match what one can experimentally carry out. This is studied more formally by using the methods of {\emph{optimal control}}, where a cost functional is associated with control function and the evolution of the system, and one uses these additional constraints to find the control function that minimizes the cost. We will consider this in more detail in Section~\ref{secOptimalControl}.

We note that even if the Kalman matrix has rank $l<n$ indicating the system is not controllable, it still possesses a controllable subspace of dimension $l$. That is, if there are $n$ initial configuration variables $x_i$, then there is a basis transformation $y_j = P_{ij}|x\rangle$, such that within the subspace spanned by the $y_1 \ldots y_l$ the system is controllable, while the subspace $y_{l+1} \ldots y_n$ is not controllable \cite{mirrahimi2004,zabczyk2020}.

With this background, we can now recognize Eq.~(\ref{eqSymplecticDETimeDependentExpectation}) as a linear control problem, with 
\begin{eqnarray}
A &=& -i\eta M \\
{\mathbf{u}(t)} &=& -i\eta {\mathbf{c}}
\end{eqnarray}
and the matrix $C$ can be constructed by noting that
$C$ is $2n\times 2n$, all off-diagonal entries are zero and $C_{kk} = 1$ if $(\eta {\mathbf{c}})_k$ is non-zero and $C_{kk} = 0$ otherwise.


Using these correspondences, we immediately we know that the system is controllable if and only if the associated Kalman matrix $K$ given by (\ref{eqKalmanMatrixDefinition}) has rank $n$. Since the matrix $M$ and vector ${\mathbf{c}}(t)$ are entirely specified by the form of the Hamiltonian, to determine the controllability we need only take the coefficients in the Hamiltonian, construct $K$, and compute the rank. Furthermore, if the system is controllable, the recipe above provides a simple way to create analytic solutions for the control functions that perform any control task required. Note that of course `controllability' here refers to expectation values as in Eq. (\ref{eqSymplecticDETimeDependentExpectation}), rather than Hilbert space controllability \cite{mirrahimi2004}. We will provide examples of the process in later Sections.

\section{Controlling the quadratic bosonic Hamiltonian}
\label{secControllingTheQBH}

The ability to understand the Hamiltonian given by Eq.~(\ref{eqFullQuadraticControlHamiltonian}) in terms of linear control theory is a powerful tool. To see why, suppose that Hamiltonian, and its evolution described by Eq.~(\ref{eqSymplecticDETimeDependentExpectation}), is controllable as determined by considering the Kalman invertibility criterion. 
Because the system is fully controllable, we know we can always find control functions ${\mathbf{c}}(t)$ such that $\langle \hat{{\boldsymbol{\alpha}}}(T) \rangle = {\mathbf{d}}$, where ${\mathbf{d}}$ is an arbitrary complex vector of our choice, regardless of the starting state of the system. Then, from Eq.~(\ref{eqAlphaExpectationSolution}), we have
\begin{equation}
-i \int_0^{t} e^{-i \eta M(T-s)} \eta {\mathbf{c}}(t) \, ds = {\mathbf{d}} - e^{-i \eta M T} \langle \hat{{\boldsymbol{\alpha}}}(0) \rangle.
\label{eqDisplacementShiftDerivation1}
\end{equation}
As $e^{-i \eta M T} \langle \hat{{\boldsymbol{\alpha}}}(0) \rangle$ is a fixed complex vector, and since we have complete freedom to choose ${\mathbf{d}}$, we see that by choosing specific control functions ${\mathbf{c}}(t)$ it is possible to ensure the quantity on the left hand side of (\ref{eqDisplacementShiftDerivation1}) can be any complex vector we desire, i.e. 
\begin{equation}
-i \int_0^{t} e^{-i \eta M(T-s)} \eta {\mathbf{c}}(t) \, ds = \boldsymbol{\beta}
\end{equation}
where $\boldsymbol{\beta}$ is arbitrary.

Now consider the solution to the full quantum problem given by Eq.~(\ref{eqFormalCauchySolution}) for the actual operators rather than expectation values. Using the control functions ${\mathbf{c}}(t)$ we obtained in the linear control case to get the quantity $\boldsymbol{\beta}$, we obtain 
\begin{eqnarray}
\hat{{\boldsymbol{\alpha}}}(t) &=& e^{-i \eta M T} \hat{{\boldsymbol{\alpha}}}(0) -i \int_0^{t} e^{-i \eta M(T-s)} \eta {\mathbf{c}}(t) \, ds \nonumber \\
&=& e^{-i \eta M T} \hat{{\boldsymbol{\alpha}}}(0) + \boldsymbol{\beta} \label{eqFullQSolution} .
\label{eqFullQSolutionShift}
\end{eqnarray}

This demonstrates that the system evolves exactly as it would do in the absence of a control (i.e. with ${\mathbf{c}}(t) = {\mathbf{0}}$), but we can perform additional arbitrary complex number shifts $\beta_j$ on the mode operators of the form 
\begin{equation}
\hat{a}_j \rightarrow \hat{a}_j + \beta_j,
\end{equation}
which is exactly the action of the displacement operator
\begin{equation}
\hat{D}(\beta) = \exp \left[ \beta \hat{a}^{\dagger} - \beta^* \hat{a}\right].
\label{eqDisplacementOperator}
\end{equation}

Crucially, we also know, from Eq.~(\ref{eqAnalyticSolutionForft}), exactly how to construct the control functions ${\mathbf{c}}(t)$ that will accomplish a specific shift $\beta$ of our choosing, at any arbitrary time $T$.

It is important to note that this is {\emph{all}} we can do. No matter what ${\mathbf{c}}(t)$ are applied, from the form of (\ref{eqFullQSolution}) all that can ever be done is adding a complex number vector to $\hat{\boldsymbol{\alpha}}$.


Furthermore, if the system is controllable, such a shift can in principle be performed arbitrarily fast, provided the coupling strengths can be arbitrarily strong. In any realistic system, of course, there will be some speed limit.

While the Kalman criterion allows the determination of whether a particular system is expectation-value controllable, it is somewhat abstract, and it is helpful to have a more physical understanding of what it means. Let the Hamiltonian (\ref{eqFullQuadraticControlHamiltonianWithLinearTerm}) have $n$ modes.
Defining the expectation values of the mode operators as ${\boldsymbol{\alpha}}(t) = \langle {\hat{\boldsymbol{\alpha}}}(t) \rangle$, the expectation value equation of motion is
\begin{equation}
\frac{d}{dt} \, {\boldsymbol{\alpha}}(t) = -i \eta M {\boldsymbol{\alpha}}(t) - i \eta {\mathbf{c}}(t),
\label{eqEOMalphaagain}
\end{equation}
where the vector ${\bf{c}}$ has $2n$ complex entries, and the matrix
\begin{equation}
M =   \begin{bmatrix}
G & B \\
B^{\ast} & G^{\ast}
\label{eqMdefinitionAgain}
\end{bmatrix}
\end{equation}
has $2n \times 2n$ complex entries. From (\ref{eqKalmanMatrixDefinition}) the associated Kalman matrix is given by 
\begin{equation}
K = -i \left[ C \,\,\,\, -i\eta M C \,\,\,\, (-i\eta M)^2 C \,\,\,\, \dots \,\,\,\, (-i\eta M)^{2n-1} C   \right]
\label{eqKalmanMatrixDefinitionForM}
\end{equation}
where $C$ constructed as a $2n\times 2n$ matrix as described in Section~\ref{secQuantumAndClassicalControlTheory}. The number of non-zero rows in $C$ is given by the number of modes we can control, i.e. the number of entries in ${\mathbf{c}}(t)$ that are non-zero. 

 
If $M$ is positive semidefinite, it is possible to diagonalize $\eta M$. The eigenvalues are real, and occur in pairs $\pm \omega_j$ where the eigenvectors are the normal modes of the system, and the $\omega_j$ are the frequencies of those normal modes \cite{blaizot1986}. Diagonalization of $\eta M$ corresponds purely to a basis change, and does not affect the controllability of the system. If the basis change is given by the unitary $V$, such that $V \eta M V^{\dagger} = D$ where $D$ is diagonal, and we define $\bar{\boldsymbol{\alpha}} = V \boldsymbol{\alpha}$ and $\overline{\eta {\mathbf{c}}} = V \eta {\mathbf{c}}$, then premultiplying(\ref{eqEOMalphaagain}) by $V$ yields
\begin{equation}
\frac{d}{dt} \, \bar{\boldsymbol{\alpha}}(t) = -i D \bar{\boldsymbol{\alpha}}(t) - i \overline{\eta \mathbf{c}}(t),
\label{eqEOMalphaagainTransformed}
\end{equation}
and in this new basis, the Kalman matrix becomes
\begin{equation}
\bar{K} = -i \left[  \bar{C} \,\,\,\, D \bar{C} \,\,\,\, D^2 \bar{C}  \,\,\,\, \dots D^{2n-1} \,\,\,\, \bar{C}   \right]
\label{eqTransformedKalman}
\end{equation}
where $\bar{C}$ is the control matrix in the new basis of normal modes.

The system is controllable if $\bar{K}$ has rank $2n$, which is equivalent to each of the $2n$ rows in (\ref{eqTransformedKalman}) being linearly independent. Clearly this will be the case if and only if all the elements making up the diagonal entries of $D$ are distinct and no row of $\bar{C}$ is entirely zero. As before, $\bar{C}$ is diagonal and $\bar{C}_{kk}$ is non-zero only if that mode is controlled in the new basis.

Since the elements of $D$ are $\pm \omega_n$, where the $\omega_n$ are the eigenvalues of the Hamiltonian, and we know that our Hamiltonian can be made positive definite, the physical interpretation of controllability corresponds to all the normal modes of the Hamiltonian having distinct energies, and the control vector $\eta {\mathbf{c}}$ having a non-zero overlap with all of the eigenvectors corresponding to the normal modes of the system.

It is difficult to go beyond this without more details of the specific Hamiltonian, so we present a physical example that is instructive in showing how controllability might be determined in a specific system. We consider a linear chain with $N$ quantum systems, such that each element in the chain is coupled only to its two nearest neighbours, by both an energy conserving hopping term and a squeezing term, and we assume we can only control the system at the end of the chain, which corresponds to a single control function, i.e. in the Hamiltonian (\ref{eqFullQuadraticControlHamiltonianWithLinearTerm}) the only coupling function that is non-zero is $c_1(t)$. We wish to know under which conditions this ensures {\emph{all}} elements of the chain are controllable.

This system is described by the Hamiltonian (\ref{eqFullQuadraticControlHamiltonianWithLinearTerm}), with the coefficients $G_{ij}$ and $B_{ij}$ given by 
\begin{equation}
G = \begin{bmatrix}
g_{11} & g_{12} & 0 & 0 & 0  \\
g_{21} & g_{22} & g_{23} & 0 & 0  \\
0 & g_{32} & g_{33} & g_{34} & 0   \\
0 & 0 & g_{43} & g_{44} & \ddots  \\
0 & 0 & 0 & \ddots & \ddots  
\end{bmatrix}, \,\,\,\,\,
B = \begin{bmatrix}
b_{11} & b_{12} & 0 & 0 & 0  \\
b_{21} & b_{22} & b_{23} & 0 & 0  \\
0 & b_{32} & b_{33} & b_{34} & 0   \\
0 & 0 & b_{43} & b_{44} & \ddots  \\
0 & 0 & 0 & \ddots & \ddots  
\end{bmatrix}
\end{equation}
where both $G$ and $B$ are $N\times N$ in size, and $\eta M$ is $2N \times 2N$ in size. The eigenvectors of $\eta M$ have $2N$ entries. We define a basis where $|n \rangle$ means a vector with a 1 in row $n$, and zeros in all the other $2N-1$ positions. In this basis, as we only have control of the first element of the chain, we have $\eta |c\rangle = c_n |1\rangle + c_{N+1} |N+1\rangle$.


We want to show that all the eigenvectors of $\eta M$ have non-zero overlap with either $|1\rangle$ or $|N+1\rangle$. We proceed using a proof by contradiction. Assume that that there exists a normalized eigenvector ${\mathbf{v}} \equiv |{\mathbf{v}}\rangle$ of $\eta M$ such that 
\begin{equation}
\langle 1| {\mathbf{v}} \rangle = \langle N+1| {\mathbf{v}} \rangle = 0,
\end{equation}
i.e. an eigenvector that has no overlap with $|1\rangle$ or $|N+1\rangle$. This means we have
\begin{equation}
\langle 1| \eta M | {\mathbf{v}} \rangle = \langle N+1| \eta M | {\mathbf{v}} \rangle = 0.
\label{eqConditionOnVOverlap}
\end{equation}

We now evaluate (\ref{eqConditionOnVOverlap}) with but now with $\eta M$ operating to the left. We have
\begin{eqnarray}
M \eta| 1 \rangle &=& g_{11} | 1 \rangle + g_{21} | 2 \rangle + b_{11}^* | N+1 \rangle   + b_{21}^* | N+2 \rangle   \\
M \eta| N+1 \rangle &=& -b_{11} | 1 \rangle - b_{21} | 2 \rangle - g_{11}^* | N+1 \rangle - g_{21}^* | N+2 \rangle.   
\end{eqnarray}
Since we know $\langle {\mathbf{v}} | 1 \rangle = \langle {\mathbf{v}} | N+1 \rangle = 0$ we have
\begin{eqnarray}
g_{21} \langle {\mathbf{v}} |  2 \rangle + b_{21}^* \langle {\mathbf{v}} | N+2 \rangle &=& 0 \\
-b_{21} \langle {\mathbf{v}} |  2 \rangle - g_{21}^* \langle {\mathbf{v}} | N+2 \rangle &=& 0 
\end{eqnarray}
This is a pair of simultaneous equations in two unknowns, so provided $|b_{21}|^2 - |g_{21}|^2 \neq 0 $  there is single unique solution given by $\langle {\mathbf{v}} |  2 \rangle = \langle {\mathbf{v}} |  N+2 \rangle =0$.


Repeating our steps, now considering $M \eta| 2 \rangle$ and $M \eta| N+2 \rangle$, we obtain a simiar condition. We find in order ensure $\langle {\mathbf{v}} |  3 \rangle = \langle {\mathbf{v}} |  N+3 \rangle =0$ we require condition $|b_{32}|^2 - |g_{32}|^2 \neq 0 $.

Continuing the process to the end of the chain, we find that ${\mathbf{v}} = {\boldsymbol{0}}$, which is contradicts our assumption that it was normalized, and therefore all eigenvectors of $\eta M$ must have some overlap with our control vector $|\eta c\rangle$, provided
\begin{equation}
|b_{ij}|^2 - |g_{ij}|^2 \neq 0, \,\,\,\, \forall i>j.
\label{eqgbupper}
\end{equation}

In order to complete the controllability analysis we need to know whether the eigenvalues of $\eta M$ are distinct. For this particular system it turns out that the best we can do is to show that the eigenvalues are {\emph{at most}} two-fold degenerate. To do this we assume that at least one eigenvalue is at least threefold degenerate, so that $\eta M$ has three eigenvectors ${\mathbf{u}}$, ${\mathbf{v}}$, ${\mathbf{w}}$ that share the same eigenvalue $\lambda$, and deduce a contradiction.

Since these three eigenvectors have the same eigenvalue, we know that any linear combination of them is also an eigenvector with
\begin{equation}
\eta M (\alpha {\mathbf{u}} + \beta {\mathbf{v}} + \gamma {\mathbf{w}}) = \lambda (\alpha {\mathbf{u}} + \beta {\mathbf{v}} + \gamma {\mathbf{w}}).
\end{equation}
We now look at the two components of each of the eigenvectors that we know cannot both be zero giving
\begin{equation}
\bar{\mathbf{u}} = \begin{bmatrix} \langle 1 | {\mathbf{u}} \rangle \\ \langle N+1 | {\mathbf{u}} \rangle \end{bmatrix}, \,\,\,\,\,
\bar{\mathbf{v}} = \begin{bmatrix} \langle 1 | {\mathbf{v}} \rangle \\ \langle N+1 | {\mathbf{v}} \rangle \end{bmatrix}, \,\,\,\,\,
\bar{\mathbf{w}} = \begin{bmatrix} \langle 1 | {\mathbf{w}} \rangle \\ \langle N+1 | {\mathbf{w}} \rangle \end{bmatrix}.
\end{equation}
$\bar{\mathbf{u}}$, $\bar{\mathbf{v}}$ and $\bar{\mathbf{w}}$ are three 2-vectors with lengths greater than zero. If we can find values for $\alpha$, $\beta$, and $\gamma$ such that 
\begin{equation}
\alpha \bar{\mathbf{u}} + \beta \bar{\mathbf{v}} + \gamma \bar{\mathbf{w}} = 0
\end{equation}
then our new eigenvector must have zeros in both position $1$ and position $N+1$, which is the contradiction we want. This is equivalent to adding three weighted 2-vectors in a plane and asking if we can get back to the origin.

Case 1: If at least two of $\bar{\mathbf{u}}$, $\bar{\mathbf{v}}$ and $\bar{\mathbf{w}}$ are parallel to each other, we give those two opposite weights and zero weight for the third.

Case 2: All three vectors are linearly independent. Since $\bar{\mathbf{u}}$ and $\bar{\mathbf{v}}$ are linearly independent, we know they span the plane, and with appropriate weights can combine to give $-\bar{\mathbf{w}}$.

Thus our contradiction is accomplished and the eigenvalues of $\eta M$ must be at most twofold degenerate, indicating even in the worst case we have controllability of a subspace corresponding to $N/2$ modes that are linear combination of the original modes. We note that such a case occurs only with a very high degree of symmetry in the Hamiltonian, and most sets coefficients $G_{ij}$ and $B_{ij}$ result in a fully controllable system.


\section{Optimal control}
\label{secOptimalControl}

In the previous sections we were concerned with expectation value control and described how, using the Kalman criterion, one could determine whether a given general multimode quadratic Hamiltonian was expectation value controllable or not. If a system was expectation value controllable, we provided an explicit method for constructing a control function that acted on combinations of position and momentum quadrature operators to move the expectation values $\langle \hat{\boldsymbol{\alpha}}(0) \rangle$ any set $\langle \hat{\boldsymbol{\alpha}}(T) \rangle$ of our choosing. Our construction provides some freedom in choosing the control function, which allows the function to be tailored to a specific system or experimental setup. This freedom, however, is purely related to the shape of the control function, and does not allow us to specify more general constraints that we may wish to enforce.

The controllability of a system in the presence of penalties and constraints is known as optimal control \cite{zabczyk2020,bechhoeffer2021}, and introduces the concept of a cost functional $J$ that provides a value for each control function. The goal is then to find the control function that minimizes that cost function. One of the most common and well-studied optimal control problems is the case where the dynamics is linear in both the controls $\mathbf{u}(t)$ and the system state variables $\mathbf{x}(t)$ (which is our situation; see Eq.~(\ref{eqSymplecticDETimeDependentExpectation})), and the cost function is quadratic, i.e.
\begin{equation}
J = \int_0 ^{T} dt \, \frac{1}{2} \left( \mathbf{x}^T Q \mathbf{x} + \mathbf{u}^T R \mathbf{u} \right),
\label{eqQuadraticCostFunction}
\end{equation}
where $Q,R$ are symmetric matrices chosen such that $R$ is semidefinite and $R$ is positive definite.

For this case it can be shown that the set of equations that must be self-consistently solved in order to minimize the cost are given by \cite{zabczyk2020,bechhoeffer2021}
\begin{eqnarray}
\dot{\mathbf{x}} &=& A \mathbf{x} + C \mathbf{u}, 
\label{eqCoupledConsistentQuadraticCostEquations1}\\
\dot{\boldsymbol{\lambda}} &=& - Q \mathbf{x} - A^T \boldsymbol{\lambda} \label{eqCoupledConsistentQuadraticCostEquations2} \\
\mathbf{u} &=& -R^{-1} C^T \boldsymbol{\lambda}.
\label{eqCoupledConsistentQuadraticCostEquations3}
\end{eqnarray}
where the state variables have been supplemented by the adjoint variables $\boldsymbol{\lambda}$ arising from the constraint of minimizing the cost $J$. In order to find analyic solutions one decouples the state and adjoint equations by introducing the matrix $S(t)$ given by
\begin{equation}
\boldsymbol{\lambda}(t) =  S(t) \mathbf{x}(t)
\end{equation}
leading to \cite{bechhoeffer2021}
\begin{equation}
\dot{S} = -Q - A^T S - SA + SCR^{-1} C^T S
\label{eqMatrixRiccati}
\end{equation}
which is a continuous time matrix Riccati equation \cite{jank2003}.

While the formulation of the problem in terms of a matrix Riccati equation allows analytic solution in a variety of cases, it is difficult to apply in the case of hard initial and final conditions, where we insist that the system start in a specific state and end in another specific state, rather than merely close to it. To see this, suppose our state $\mathbf{x}(t)$ consists of $n$ state variables, so that $\mathbf{\lambda}(t)$ also has $n$ components. Eqs.~(\ref{eqCoupledConsistentQuadraticCostEquations1} -- \ref{eqCoupledConsistentQuadraticCostEquations3}) consist of $2n$ coupled first order differential equations, and if we specify $\mathbf{x}(0)$ and $\mathbf{x}(T)$ we have $2n$ boundary conditions and the problem is well-specified. If we use Eq.~(\ref{eqMatrixRiccati}), however, we cannot specify initial conditions, since we do not know what the initial and final boundary conditions for $\boldsymbol{\lambda}$ must be.

With this background, we now provide an example demonstrating how our symplectic formulation of the quadratic bosonic Hamiltonian allows not just for expectation value control, but also optimal expectation value control that can minimize costs such as total energy or total control effort.

To begin, we note that if we work in the $\hat{x},\hat{p}$ basis rather than the $\hat{a},\hat{a}^\dagger$ basis, all our quantities are real, and more physically intuitive. In this basis, following the methods of Section~\ref{secQHOandNotation}, if our Hamiltonian is given by
\begin{eqnarray}
\hat{H} &=& \sum_{i,j} \left( G^x_{ij} \hat{x}_i \hat{x}_j +  G^p_{ij} \hat{p}_i \hat{p}_j \right) + B_{ij} \left( \hat{x}_i \hat{p}_j + \hat{p}_i \hat{x}_j \right) \nonumber \\
&& + \sum_i \left( c_i^x(t) \hat{x}_i + c_i^p(t) \hat{p}_i \right) 
\label{eqFullHxp}
\end{eqnarray}
we find
\begin{equation}
\frac{d}{dt} \, \langle \hat{\mathbf{x}} \rangle = A \langle \hat{\mathbf{x}} \rangle + {\mathbf{c}}(t) 
\end{equation}
where
\begin{eqnarray}
A &=& 2 \begin{bmatrix} B & G^p \\ -G^x & -B \end{bmatrix} \\
\hat{\mathbf{x}} &=& \begin{bmatrix} \hat{x}_1 & \hat{x}_2 & \hat{x}_3 & \ldots & \hat{p}_1 & \hat{p}_2 & \hat{p}_3 & \ldots \end{bmatrix}^T \\
{\mathbf{c}}(t) &=& \begin{bmatrix} c^p_1(t) & c^p_2(t) & \ldots -c^x_1(t) & -c^x_2(t) & \ldots  \end{bmatrix}^T.
\end{eqnarray}

For our example we take a linear chain of oscillators with nearest neighbour couplings as discussed in Section~\ref{secControllingTheQBH}, and consider the two element case where we can control the position and momentum of the first element with some time-dependent function $u(t)$. An example of such a Hamiltonian is given by
\begin{eqnarray}
\hat{H} &=& 2\hat{x}_1^2 + 2\hat{x}_2^2 +  2\hat{p}_1^2 +  2\hat{p}_2^2 + \hat{x}_1 \hat{x}_2 + \hat{x}_2 \hat{x}_1 \nonumber \\
&& + \hat{x}_1 \hat{p}_2 + \hat{p}_2 \hat{x}_1 + u(t) (\hat{x}_1 + \hat{p}_1)
\label{eqTwoModeExampleHamiltonian}
\end{eqnarray}
giving
\begin{eqnarray}
A &=& 2\begin{bmatrix} 0 & 1 & 2 & 0 \\ 1 & 0 & 0 & 2 \\ -2 & -1 & 0 & -1  \\ -1 & 2 & 1 & 0  \end{bmatrix} \nonumber \\
\mathbf{x} &=& \begin{bmatrix} \langle\hat{x}_1\rangle & \langle\hat{x}_2\rangle & \langle\hat{p}_1\rangle & \langle\hat{p}_2\rangle \end{bmatrix}^T \nonumber \\
{\mathbf{c}} &=& \begin{bmatrix} 1 & 0 & -1 & 0 \end{bmatrix}^T
\label{eqTwoModeSimpleExampleControl}
\end{eqnarray}
where have used the notation $\mathbf{x} = \langle \hat{\mathbf{x}} \rangle$. In order to determine the controllability of the system we note that since ${\mathbf{u}}(t) = [u(t), 0, -u(t), 0]$ we have $C=\text{diag}\{ 1,0,1,0\}$ and the associated Kalman constructed via 
\begin{equation}
K = \left[ C \,\,\,\, A C \,\,\,\, A^2 C \,\,\,\, \dots \,\,\,\, A^{2n} C   \right]
\end{equation}
has rank $4$, indicating the system is controllable.

We choose $Q=q \mathbb{1}$ and since we have a single control function we must have $R = r$ where $r$ is a scalar, giving a cost function
\begin{equation}
J = \int_0 ^{T} dt \, \frac{1}{2} \left( q|\mathbf{x}|^2 + r |u|^2 \right).
\label{eqSpecificQuadraticCostFunction}
\end{equation}

\begin{figure}[htp]
\subfloat{
  \includegraphics[width=0.95\columnwidth]{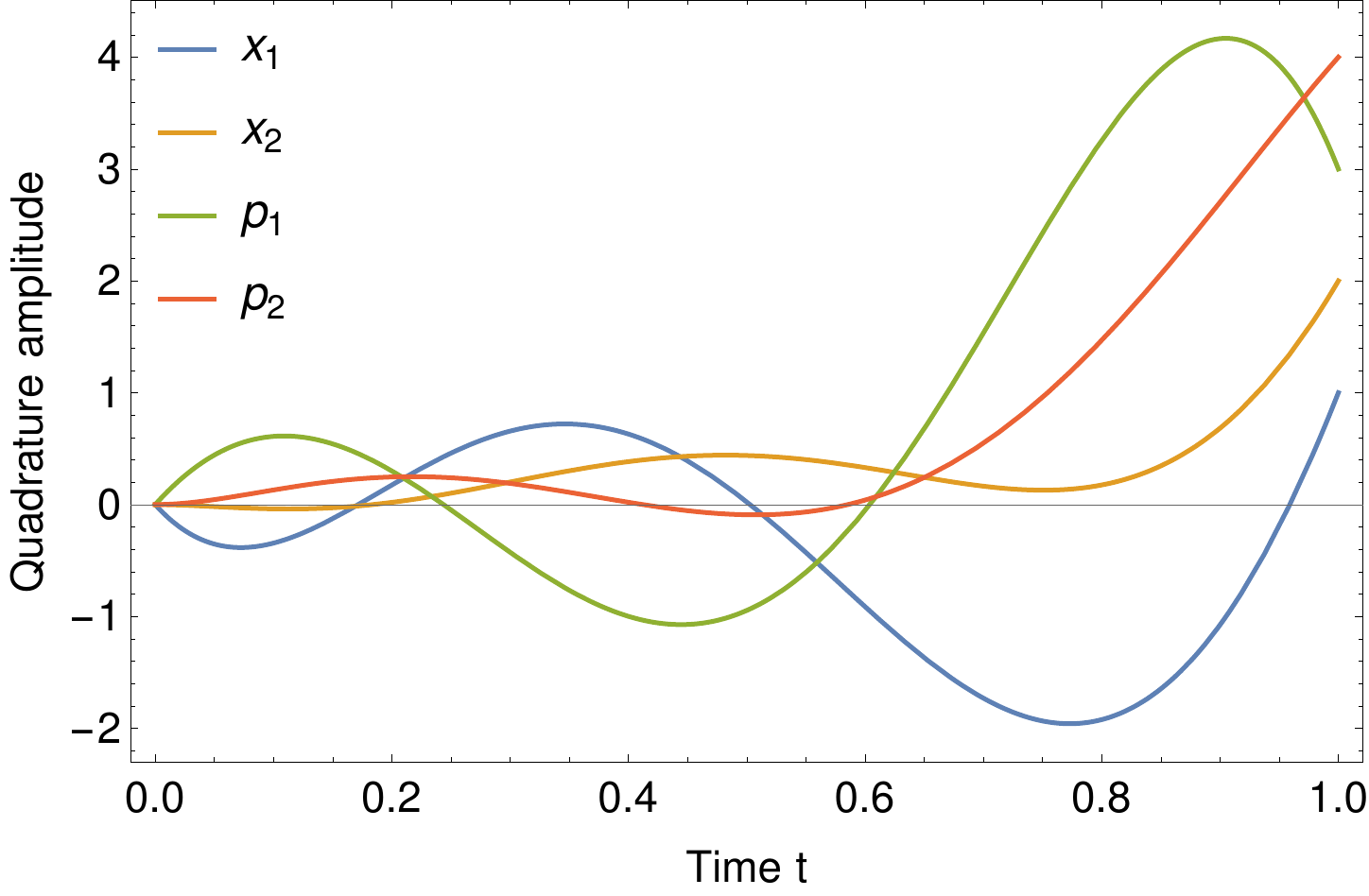}
}

\subfloat{
  \includegraphics[width=0.95\columnwidth]{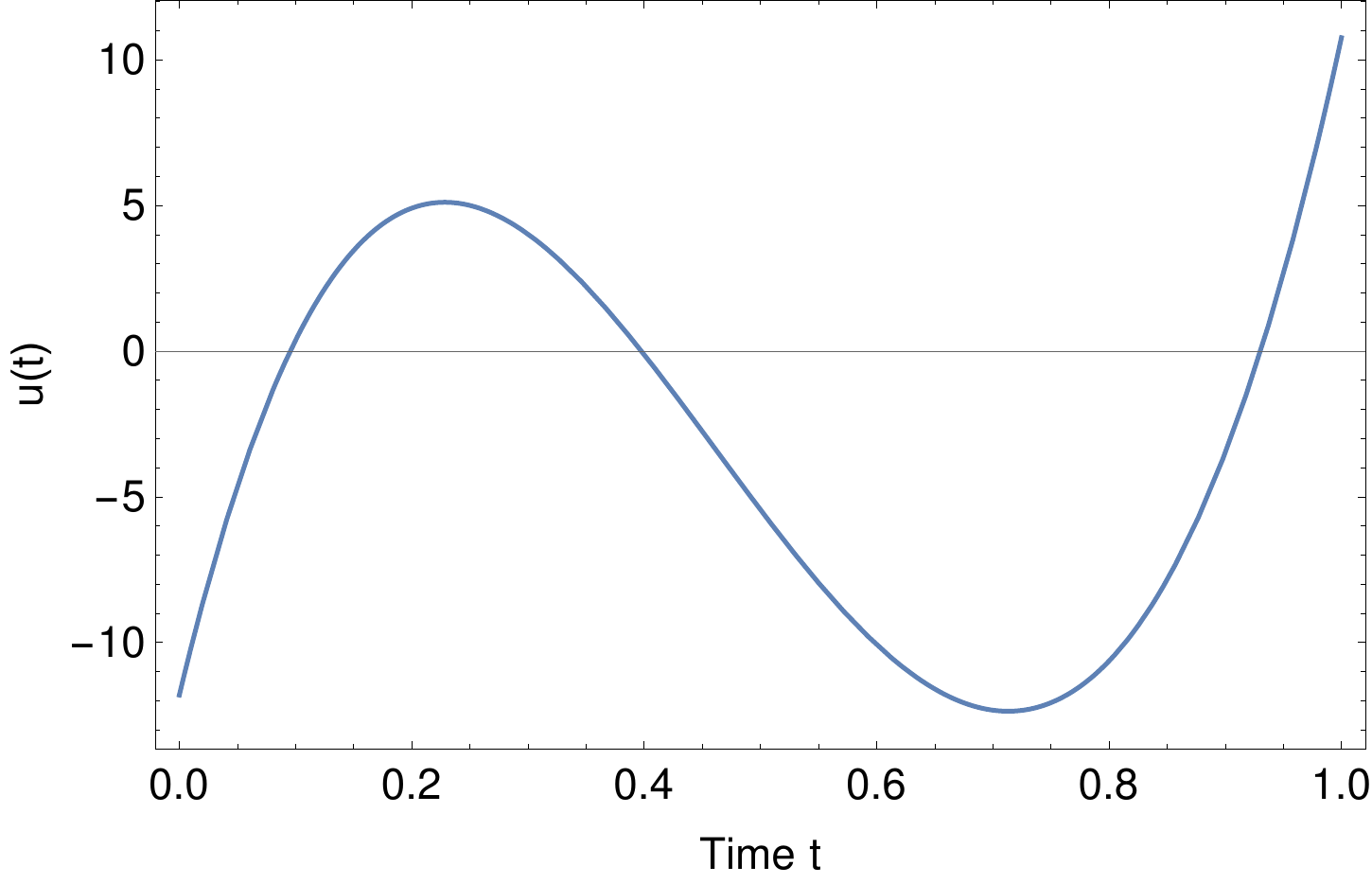}
}
\caption{The solution to an optimal control problem for the Hamiltonian (\ref{eqTwoModeExampleHamiltonian}) that minimizes the cost (\ref{eqSpecificQuadraticCostFunction}) for $q=r=1$, and with initial and final conditions chosen to be $\mathbf{x}(0) = (0,0,0,0)$ and $\mathbf{x}(1) = (1,2,3,4)$. Top panel: Evolution of the state variables. Bottom panel: the control function $u(t)$ that accomplishes this.}
\label{fig:SimpleExampleControl}
\end{figure}

 We can now specify initial and final conditions and solve (\ref{eqCoupledConsistentQuadraticCostEquations1} -- \ref{eqCoupledConsistentQuadraticCostEquations3}) numerically to obtain the system dynamics and the control function that minimizes $J$. Figure~\ref{fig:SimpleExampleControl} showns the results for the case $q=r=1$, and with initial and final conditions chosen to be $\mathbf{x}(0) = (0,0,0,0)$ and $\mathbf{x}(1) = (1,2,3,4)$.
 

\begin{figure}[htp]
\includegraphics[width=0.95\columnwidth]{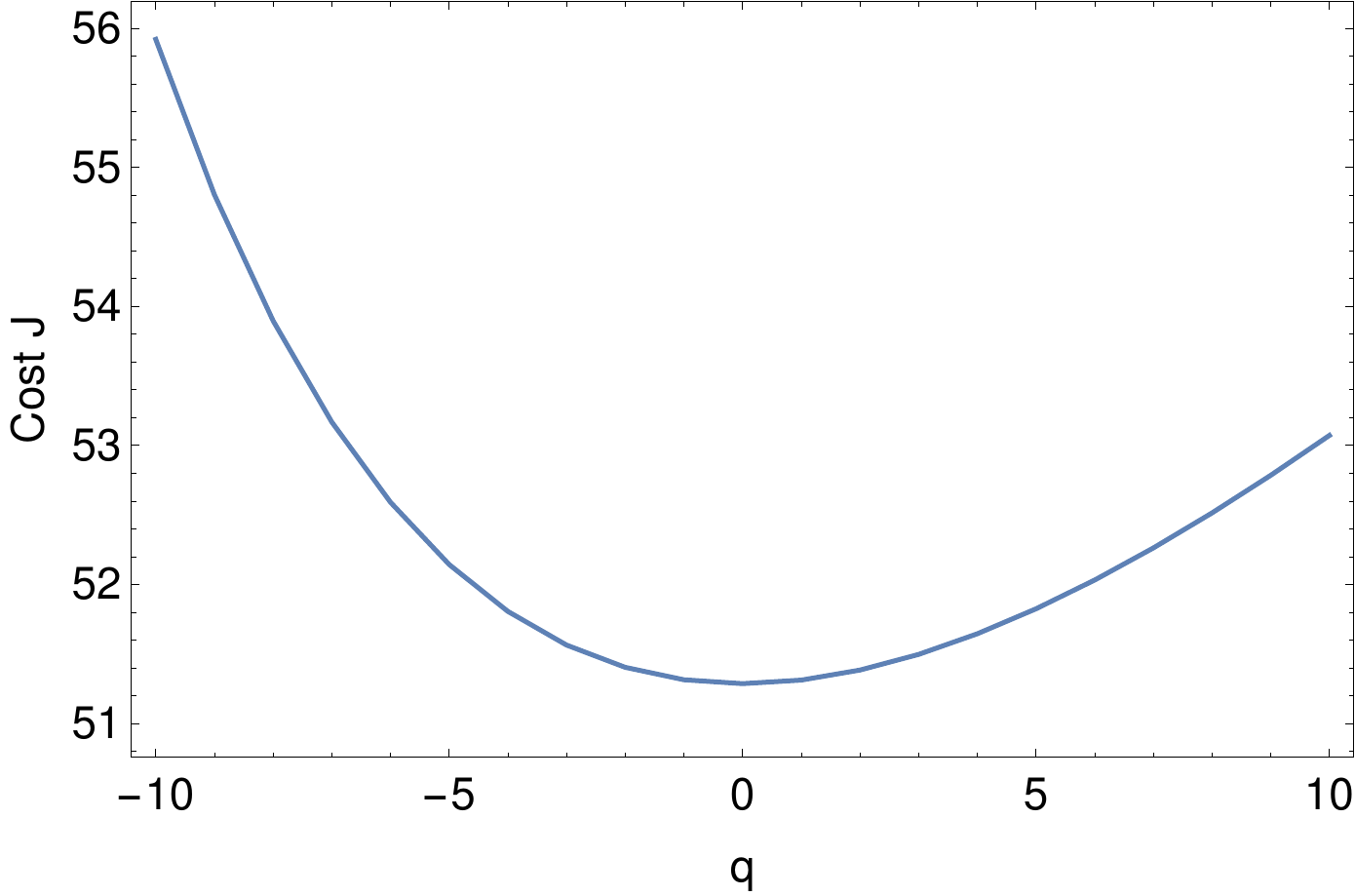}
\caption{The minimal possible cost for moving the system described by the Hamiltonian (\ref{eqTwoModeExampleHamiltonian}) from $\mathbf{x}(0) = (0,0,0,0)$ to $\mathbf{x}(1) = (1,2,3,4)$ for a range of control and displacement penalties. The cost is calculated according to the cost function (\ref{eqSpecificQuadraticCostFunction}), with $r=1$ fixed and $q$ allowed to vary.}
\label{fig:SimpleExampleCostsAsWeVaryq}
\end{figure}

We can also see how the total cost $J$ varies as we change $q$ and $r$ depending on whether we prioritize keeping total energy low ($q/r \ll 1$) or using minimal effort ($q/r \gg 1)$. This tradeoff is shown in Figure \ref{fig:SimpleExampleCostsAsWeVaryq}. 

We note that in the restricted case of minimal effort control ($q=0$) there is an analytic solution to this optimal constrol problem \cite{zabczyk2020}, with the minimal cost control function given by
\begin{equation}
{\mathbf{u}}(t) = -C^\dagger \exp[A^\dagger(T-t)] Q_T^{-1} (\exp[AT]\mathbf{x}(0) - \mathbf{x}(T)) 
\label{eqOptimalControlSolutionZabczykMinimumEffort}
\end{equation}
 where $Q_T$ is the controllability Grammian given by
\begin{equation}
Q_T = \int_0^T e^{As} C C^\dagger e^{A^\dagger s} \, ds.
\label{eqControllabilityGrammian}
\end{equation}
The cost of this solution is given by 
\begin{equation}
\int_0^T |u(t)|^2 dt = \langle Q_T^{-1}\left( e^{AT} \mathbf{x}(0) - \mathbf{x}(T)\right), e^{AT} \mathbf{x}(0) - \mathbf{x}(T) \rangle
\end{equation}
which agrees with the minimum in Figure~\ref{fig:SimpleExampleCostsAsWeVaryq}.

\section{Applications}
\label{secApplications}

Harmonic oscillators are ubiquitous in physics, partly because they describe the quantized electromagnetic field, and partly because they are applicable to any bound quantum system with a confining potential since near the minimum the potential will be approximately harmonic. The full quadratic bosonic oscillator described by Eq.~(\ref{eqFullQuadraticControlHamiltonian}) is even more general as it includes squeezing terms as well as the most general type of linear coupling. This generality ensures such a Hamiltonian describes a very large range of quantum systems, including electromagnetic resonators, micromechanical systems, phononic systems, quantum field theories and many others. In this section we consider a small illustrative sample of such systems, and consider how our results on controllability are relevant in each case.

\subsection{Wavepacket transport beyond the adiabatic limit}

High-fidelity quantum transfer of quantum states from one location to another is an essential element of quantum information processing. If massive particles rather than photons are used, this involves the use of trapping potentials such as time-averaged rf fields, magneto-optical traps, optical lattices or optical levitation, and all such traps are locally harmonic near the minimum. Once the particle is trapped, regardless of whether it is an atom, a molecule or a macroscopic object, it is often the case that it must be moved without disturbing the quantum state of the system. This can be accomplished by moving the trapping potential arbitrarily slowly, but this has clear drawbacks in terms of quantum information processing speed, noise and decoherence. For that reason fast transport far from the adiabatic limit is an area of intense interest \cite{calarco2009, lam2021}.

Most transport schemes involve moving the position of the trap minimum as a function of time, with the transport path and velocity chosen such that when the transport comes to a halt, the wavefunction of the trapped quantum system is identical to the state immediately before the transport began, usually the ground state. When transporting far from the adiabatic limit the wavefunction will undergo significant distortion during the transfer process, but ideally the path and velocity is chosen such that any excitations or heating should be cancelled by the final position.

Such nonadiabatic transfer paths can be found through quantum control optimization \cite{lam2021}, machine learning, or a single specific solution \cite{calarco2009}, and often do not provide insight into {\emph{why}} that particular path might be a good one, or allow the generation of an arbitrary number of alternative paths with properties that may be more experimentally suitable, or allow the imposition of cost functions. Our formulation of the problem in terms of a symplectic linear control problem provides all of these. 

We analyse the system for a single mode, as this is the situation generally considered, and also because it is the simplest possible application of our theory and serves as a good example. We note however, that extension to an abitrary number of harmonic oscillator modes corresponding to other degrees of freedom of the trap, as well as arbitrary cross-mode couplings is trivial. 

The standard formulation of the problem is to consider a Hamiltonian of the form
\begin{equation}
\hat{H} = \frac{\hat{p}^2}{2m} +\frac{1}{2} m \omega^2 \left( \hat{x}  + u(t) \right)^2
\label{eqSingleModeWavepacketTransportHamiltonian}
\end{equation}
where $u(t)$ is a control function that moves the position of the trap minimum.

In this form the Hamiltonian is clearly equivalent to our quadratic bosonic Hamiltonian (\ref{eqFullHxp}), with only a single mode and no squeezing operations, and we have 
\begin{equation}
G^x = \frac{1}{2} m \omega^2, \,\,\,\,\,  G^p = \frac{1}{2m}, \,\,\,\,\, B=0, \,\,\,\,\, {\boldsymbol{x}} = \begin{bmatrix} \langle \hat{x} \rangle \\ \langle \hat{p} \rangle \end{bmatrix},
\end{equation}
\begin{equation}
A = \begin{bmatrix} 0 & \frac{1}{m} \\ -m\omega^2 & 0 \end{bmatrix},  \,\,\,\,\, {\bf{c}}(t) =   \begin{bmatrix} 0 \\ -m\omega^2 u(t) \end{bmatrix}, \,\,\,\,\, C = \begin{bmatrix} 0 & 0 \\ 0 & 1  \end{bmatrix} 
\end{equation}
\begin{equation}
\frac{d {\mathbf{x}}}{dt} = A {\mathbf{x}} + C {\bf{c}}(t)
\end{equation}
Choosing $\omega = 1$ and $m=1$, and utilizing our definition for the Kalman matrix (\ref{eqKalmanMatrixDefinitionForM}) we obtain
\begin{equation}
K = \begin{bmatrix} 0 & 0 & 0 & 1 \\ 0 & 1 & 0 & 0 \end{bmatrix}.
\end{equation}
$K$ has rank 2, which confirms the system is controllable. This means we can find a $u(t)$ such that we perform the displacement operation (\ref{eqDisplacementOperator}) with a $\beta$ of our choosing, over a time period $T$, also of our choosing. From Eq.~(\ref{eqDisplacementOperator}), if $\beta$ is chosen real, we perform a perfect fidelity transport of our quantum state $|\psi(x,t)\rangle \rightarrow |\psi(x+\beta, t+T)\rangle$.

We have a large amount of freedom in choosing $u(t)$ --- we can choose from the classes of function $\mu(t)$ described in Section~\ref{secQuantumAndClassicalControlTheory}, or we can choose a cost function and solve the system of equations (\ref{eqCoupledConsistentQuadraticCostEquations1}) -- (\ref{eqCoupledConsistentQuadraticCostEquations3}).
Of course, we are not limited purely to displacements of the intitial state --- we can also clearly provide any arbitrary centre of mass momentum shift that we choose.

\begin{figure}[htp]
\includegraphics[width=0.95\columnwidth]{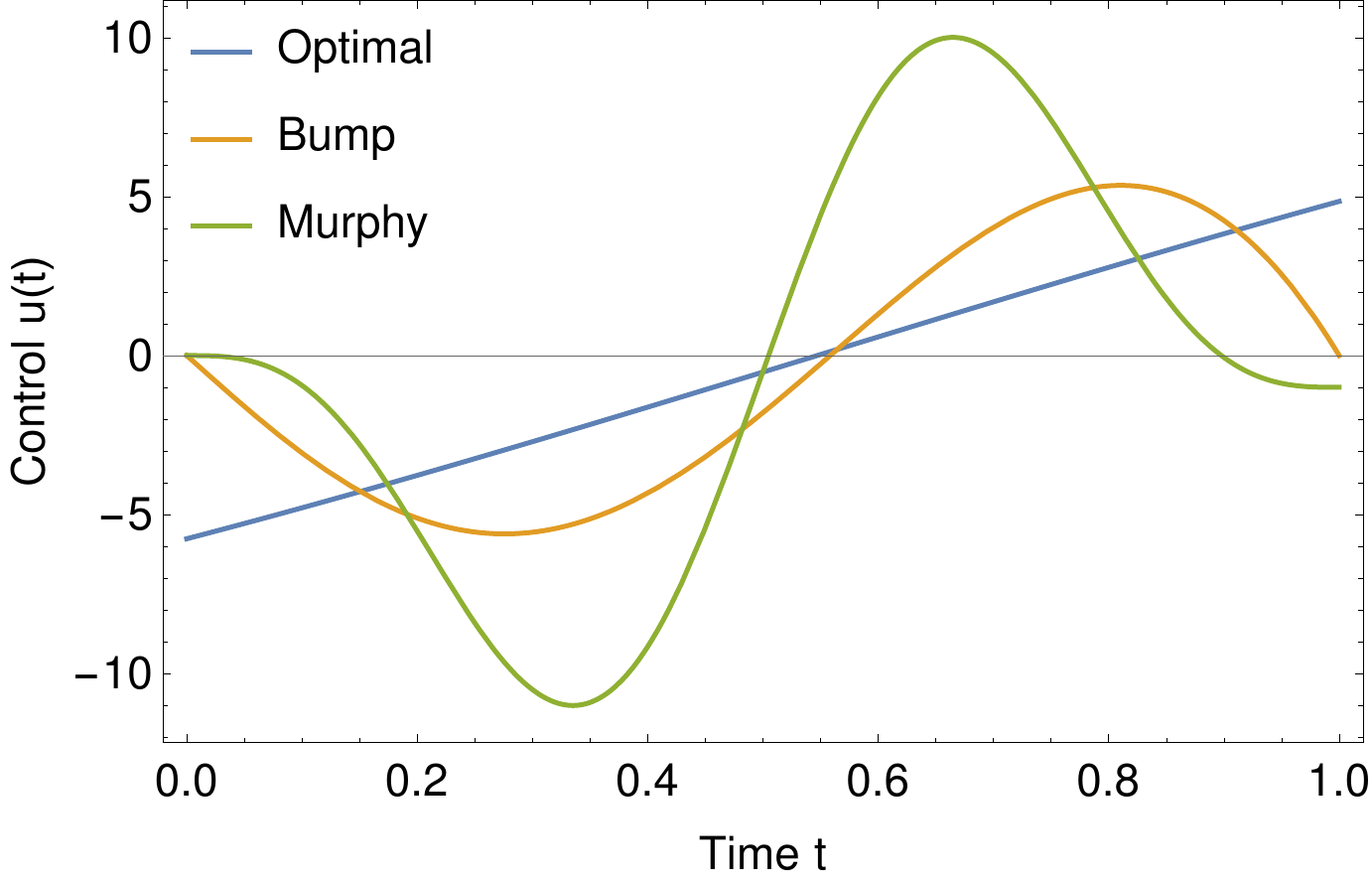}
\caption{Three control functions $u(t)$ that perform perfect fidelity wave function transport from $|\psi(x,t=0) \rangle $ to $|\psi(x-1,t=0) \rangle$ using the Hamiltonian (\ref{eqSingleModeWavepacketTransportHamiltonian}). At the final time the wavepacket has exactly the same functional form as at $t=0$, but displaced by one unit. The orange line is a $u(t)$ is computed using the method in Section~\ref{secQuantumAndClassicalControlTheory} with a standard bump function which takes no account of cost; the blue line is a $u(t)$ determined using the method described in Section~\ref{secOptimalControl} which respects a minimum effort cost function; and the green line is the explicit transport function given by Murphy {\emph{et al}}. \cite{calarco2009}. The costs of the three methods are  $J=15.3$, $J=9.97$ and $J=40.4$ respectively.}
\label{fig:CompareVariousControlFunctions}
\end{figure}

Suppose we wish to move our system from $\langle \hat{x}(0) \rangle = \langle \hat{p}(0) \rangle = 0$ to $\langle \hat{x}(1) \rangle =1 $, $\langle \hat{p}(1) \rangle = 0$, and we wish to minimize the cost of the control, i.e. our figure of merit is minimizing the cost function 
\begin{equation}
J = \int_0^T |{\mathbf{u}}(t)|^2 \, dt.
\end{equation}
We compute the control function $u(t)$ using three different methods: 1) The explicit method outlined in Section~\ref{secQuantumAndClassicalControlTheory} with $\mu(t) = t^2(1-t^2)$ which takes no account of the cost function, 2) the method described in Section~\ref{secOptimalControl} using Eqs.~(\ref{eqOptimalControlSolutionZabczykMinimumEffort}) and (\ref{eqControllabilityGrammian}) which respects the cost function, and 3) the explicit transport function given by Murphy {\emph{et al}}. \cite{calarco2009}, and plot them in Figure~\ref{fig:CompareVariousControlFunctions}. All these control functions perform perfect fidelity transport, but if we calculate the cost $J = \int_0^1 |u(t)|^2 dt$ we find costs of $J=15.3$, $J=9.97$ and $J=40.4$ respectively, with the optimal control method providing the lowest cost as expected.

\subsection{Implementation of an echoed conditional displacement gate}

Recently there has been a great deal of interest in techniques to amplify a weak dispersive interaction between a quantum system and a cavity mode, represented by a bosonic harmonic oscillator. This is relevant for systems as diverse as micromechanical oscillators \cite{arriola2019}, superconducting qubits \cite{blais2021}, optomechanics \cite{aspelmeyer2014}, quantum accoustics \cite{stazinger2018,chu2017}, and semiconductor-based qubits \cite{burkhard2020}. The essential part of such a technique is to first displace the state of the oscillator into a coherent state far from the vacuum, and then allow it to interact with the system of interest, using the enhanced "lever arm" of the displacement to enhance the interaction, before displacing the cavity back to its desired quadrature state. 

One example of such a technique is a scheme to couple a cavity mode to a superconducting transmon qubit and use displacements of the cavity field to amplify the interaction between the cavity mode and the qubit, allowing, for example, QND measurement of the qubit state \cite{touzard2019} or the generation of an echoed displacement gate (ECD) \cite{eickbusch2022}. We use the latter to illustrate the usefulness of our technique, but our method is more generally applicable to any such scheme to enhance the weak dispersive coupling of a bosonic mode.

The ECD gate is an entangling gate of the form
\begin{equation}
{\text{ECD}}(\beta) = D(\beta/2) |e\rangle \langle g| + D(-\beta/2) |g\rangle \langle e |
\label{eqECDdefinition}
\end{equation}
where $D(\beta) = e^{\beta a^\dagger - \beta^* a}$ is the displacement operator of a bosonic mode $a$, and $|g\rangle$ and $|e\rangle$ are the states of a two-level system, for example qubit states such superconducting transmons. Along with local operations this gate is universal \cite{eickbusch2022}.

The Hamiltonian for the dispersive interaction is 
\begin{equation}
H = \frac{\chi}{2} a^{\dagger} a \sigma_z + u(t) a^\dagger + u^* (t) a
\label{eqDispersiveHWithDrive}
\end{equation}
where $\chi$ is the strength of the dispersive coupling and $u(t)$ is the amplitude of the time-dependent drive. Eq.~(\ref{eqDispersiveHWithDrive}) is often analysed in a time-dependent displaced frame giving the Hamiltonian
\begin{equation}
H = \frac{\chi}{2} a^{\dagger} a \sigma_z + \frac{\chi}{2} \left( \alpha(t) a^\dagger + \alpha^* (t) a \right) \sigma_z + \frac{\chi}{2} |\alpha(t)|^2 \sigma_z
\label{eqDispersiveHWithDriveDisplacedFrame}
\end{equation} 
where $\alpha(t)$ is given by solving $\partial_t \alpha(t) = -i u(t) - \kappa \alpha(t)/2$ with $\kappa$ the amplitude damping rate, and describes the classical response to a resonant drive. In this picture, the ECD gate is accomplished in a time $T$ by beginning with the cavity vacuum state and performing the displacement $D(\alpha)$. The system is then left to evolve for a time $T/2$, followed by the operation $D(-\alpha \cos(\chi T/4)) X D(-\alpha \cos(\chi T/4))$, where $X$ is a $\pi$ pulse corresponding to an $X$ gate, and then allowed to evolve for another time $T/2$. A final displacement $D(\alpha \cos(\chi T/2))$ results in the gate (\ref{eqECDdefinition}). With an intermediate oscillator displacement of $\alpha_0$, the gate is accomplished in a time of order $T\sim |\beta|/(\chi \alpha_0)$.

We will now demonstrate how the ECD gate can be accomplished in a similar time using our approach, obtaining families of functions that perform the control perfectly, while still allowing flexibility in the form of the drive amplitude $u(t)$. This means  our scheme can produce time-dependent drives that are continuous rather than the theoretically easier but experimentally hard-to-implement delta functions or piecewise constant functions generated by approaches like GRAPE \cite{khaneja2005,machnes2011,floethe2012}. 

 We work with (\ref{eqDispersiveHWithDrive}), and initially assume the qubit is in an eigenstate of $\sigma_z$, with $\sigma_z|g\rangle = \frac{1}{2}|g\rangle$ and $\sigma_z|e\rangle = -\frac{1}{2}|e\rangle$. Considering only the expectation value of the cavity state $\langle a \rangle$ we obtain two possible equations of motion depending on whether the eigenvalue is positive or negative, given by
\begin{eqnarray}
\frac{d}{dt} \langle a \rangle^+ &=& -i \frac{\chi}{2} \langle a \rangle^+ - i u(t) \label{eqECD_EOM1} \\
\frac{d}{dt} \langle a \rangle^- &=& i \frac{\chi}{2} \langle a \rangle^-- i u(t). \label{eqECD_EOM2}
\end{eqnarray}
There are two additional equations for $\langle a^\dagger \rangle^+$ and $\langle a^\dagger \rangle^-$, but they are merely complex conjugates of Eqs.~(\ref{eqECD_EOM1}) and (\ref{eqECD_EOM2}) and are redundant so we ignore them. We begin with the cavity in a vacuum, so $\langle a(0) \rangle^{\pm} = 0$, and wish to find a control function $u(t)$ such that 
\begin{equation}
\langle a(T) \rangle^+ = \frac{\beta}{2}, \,\,\,\,\,\, \langle a(T) \rangle^- = -\frac{\beta}{2}.
\label{eqECDFinalConditions}
\end{equation}

\begin{figure}[htp]
\includegraphics[width=0.95\columnwidth]{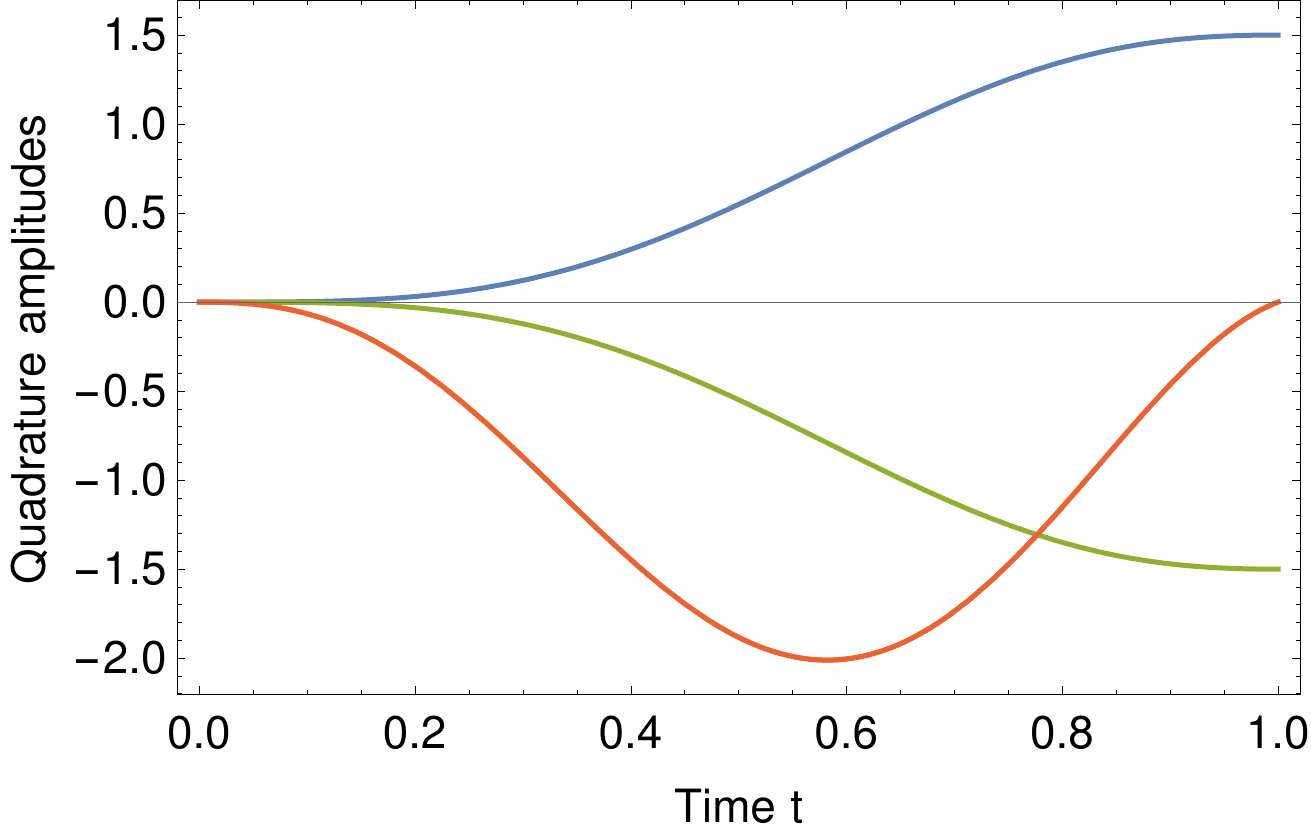}
\caption{Illustration of a control pulse displacing a cavity vacuum field to $\beta/2$ or $-\beta/2$ depending on whether the coupled qubit is in $|g\rangle$ or $|e\rangle$. The plot shows the real and imaginary parts of quadrature amplitudes resulting from the solution of (\ref{eqECD_EOM1}) and (\ref{eqECD_EOM2}) using the time dependent drive given by (\ref{eqTimeDependentControlForECD}). The blue line shows ${\text{Re}} (\langle a(t) \rangle^+)$ when the qubit is in state $|g\rangle$ and the green line shows ${\text{Re}} (\langle a(t) \rangle^-)$ when the  qubit state is $|e\rangle$. The imaginary parts of both $\langle a(t) \rangle^{\pm}$ are identical and shown in red. Parameters: $T=1$, $\beta=3$, $\chi = 3$. We see that the control function successfully displaces the cavity to $\pm \beta$ depending on the state of the qubit.}
\label{fig:ECDQuadratures}
\end{figure}

This is exactly the type of linear control problem described in Sections~\ref{secQuantumAndClassicalControlTheory} and \ref{secControllingTheQBH}. Specifically, we have a linear control problem of the form (\ref{eqLinearControlDefinition}), identifying
\begin{equation}
A = i \, \frac{\chi}{2} \begin{pmatrix} 1 & 0 \\ 0 & -1 \end{pmatrix}, \,\,\,\,\, C = \begin{pmatrix} 1 \\ 1 \end{pmatrix}.
\label{eqECD_AandCDefinition}
\end{equation}
Using the final conditions given by (\ref{eqECDFinalConditions}) to define the goal condition ${\mathbf{g}} = [\beta/2, -\beta/2 ]^T$, and employing Eqs.~(\ref{eqBumpFunctionDefinition}),  (\ref{eqroftdefinition}), and (\ref{eqAnalyticSolutionForft}) we obtain an explicit control function
\begin{eqnarray}
u(t) &=& 30 t (t-1) \left[ (2-4t) \cos \left( \frac{3(t-1)}{2} \right) \right. \nonumber \\ 
&& \,\,\,\,\, \left. + 3t(t-1) \sin \left( \frac{3(t-1)}{2} \right) \right].
\label{eqTimeDependentControlForECD}
\end{eqnarray}
The result of such a control pulse is a unitary $U_{ECD}$ that acts as follows:
\begin{equation}
U_{ECD} |0,g \rangle  = |\frac{\beta}{2}, g\rangle, \,\,\,\,\, U_{ECD} |0,e \rangle  = |-\frac{\beta}{2}, e\rangle.
\end{equation}
With the application of a final fast local $X$ gate, one obtains the gate (\ref{eqECDdefinition}) as desired. The action of the control pulse (\ref{eqTimeDependentControlForECD}) is shown in Figure~\ref{fig:ECDQuadratures}, demonstrating how the cavity state is displaced to either $\pm \beta$ depending on the qubit state. We note that this approach is of course also applicable to the displaced frame Hamiltonian given by Eq.~(\ref{eqDispersiveHWithDriveDisplacedFrame}), with the same matrix $A$ as in (\ref{eqECD_AandCDefinition}), but with $C=[1, -1]^T$. 

In addition, for this system the determinant of the Kalman matrix $K$ (\ref{eqKalmanMatrixDefinition}) is $\text{det} \, K = -i \chi$ and is non-zero, meaning $K$ has full rank, indicating that the system is controllable. This shows that we are not merely restricted to producing pair of cavity states such as $| \beta, g\rangle, |-\beta, e\rangle$ conditioned on the state of the qubit, but can just as easily produce any arbitrary result of the form $| \beta_1, g\rangle, |\beta_2, e\rangle$ for any $\beta \in {\mathbb{C}}$.

\subsection{Optomechanics}

The field of optomechanics involves coupling electromagnetic fields to micromechanical resonators \cite{aspelmeyer2014}. As quantized electromagnetic fields and phonons are both described by bosonic operators, particularly in the cavity regime, this subject is directly applicable to our theory. 

The simplest optomechanical system involves an electromagnetic cavity mode coupled to a change in length of the cavity, for example a moveable mirror at one end. The standard optomechanical Hamiltonian is given by
\begin{equation}
\hat{H} = -\hbar \Delta \hat{a}^{\dagger} \hat{a} + \hbar \Omega \hat{b}^{\dagger} \hat{b} - \hbar g \hat{a}^{\dagger} \hat{a} (\hat{b} + \hat{b}^{\dagger})
\end{equation}
where $\hat{b} + \hat{b}^{\dagger}$ describes the displacement of the mirror. The obvious generalization is to allow the cavity mode to many mechanical modes, while also allowing these mechanical modes couple to each other. In this situation the Hamiltonian becomes
\begin{eqnarray}
\hat{H}/\hbar &=& -\Delta \hat{a}^{\dagger} \hat{a} + \sum_{i,j} \left( G_{ij} \hat{b}^{\dagger}_i \hat{b}_j + \frac{1}{2} B_{ij} \hat{b}^{\dagger}_i \hat{b}^{\dagger}_j +  \frac{1}{2} B^{\ast}_{ij} \hat{b}_j \hat{b}_i \right) \nonumber \\
&& \,\,\,\, -  \hat{a}^{\dagger} \hat{a} \sum_i g_i \left( \hat{b}_i + \hat{b}^{\dagger}_i \right)
\label{eqOptomechanicsH2}
\end{eqnarray}
where we have allowed the possibility of squeezing terms between the mechanical modes, which can occur, for example with pulsed optomechanical interactions \cite{bennett2018} or intra-cavity parametric amplication \cite{agarwal2016}.

Using the formalism in Section~\ref{secQHOandNotation}, Eq.~(\ref{eqOptomechanicsH2}) can be written as
\begin{equation}
\hat{H} = \frac{1}{2} \hat{{\boldsymbol{\beta}}}^{\dagger} M \hat{{\boldsymbol{\beta}}} -\frac{1}{2} {\text{Tr}}(G) - \hat{a}^{\dagger} \hat{a} {\boldsymbol{g}}^{\dagger} \hat{{\boldsymbol{\beta}}} -\Delta \hat{a}^{\dagger} \hat{a}
\end{equation}
The cavity operator $\hat{a}$ commutes with the mechanical mode operators $\hat{b}_i$, so that when we take commutators to determine the equations of motion for the $\hat{b}_i$ we obtain
\begin{equation}
i \hbar \frac{d}{dt} \, \hat{{\boldsymbol{\beta}}} = 
\begin{bmatrix}  
G & B \\
-B^{\ast} &-G^{\ast}
\end{bmatrix} 
\hat{{\boldsymbol{\beta}}} + 
\hat{a}^{\dagger} \hat{a} \begin{bmatrix}  
g_{i}  \\
-g_{i}^*
\end{bmatrix},
\end{equation}
where 
\begin{eqnarray}
\hat{{\boldsymbol{\beta}}} &=&  
\begin{bmatrix}
\hat{b}_1 &
\hat{b}_2 &
\hat{b}_3 &
\cdots &&
\hat{b}^{\dagger}_1 &
\hat{b}^{\dagger}_2 &
\hat{b}^{\dagger}_3 &
\cdots
\end{bmatrix}^T, \\
{\mathbf{g}} &=&  
\begin{bmatrix}
g_1 &
g_2 &
g_3 &
\cdots &
g_1 &
g_2 &
g_3 &
\cdots
\end{bmatrix}^T, 
\end{eqnarray}
since the $g_i$ are real. The associated equations of motion for the expectation values that govern linear controllability are given by 
\begin{equation}
i \hbar \frac{d}{dt} \, \langle \hat{{\boldsymbol{\beta}}} \rangle = \eta M \langle \hat{{\boldsymbol{\beta}}} \rangle - \langle \hat{a}^{\dagger} (t) \hat{a}(t) \rangle \eta {\boldsymbol{g}}.
\label{eqOptomechanicsExpectationEOM}
\end{equation}

When considering a single mode cavity driven by a laser, standard input/output formalism shows the electromagnetic field inside the cavity is a coherent state. For a strong drive, back action on the electromagnetic field by the mechanical system is negligible, allowing the cavity mode operator $\hat{a}(t)$ to be described by the complex number $\alpha(t)$. The magnitude of $\alpha(t)$ is related to the number photons in the cavity $N$ via $N=|\alpha(t)|^2$, which, in turn, is given by the laser power, and is an easily controllable parameter. Eq.~(\ref{eqOptomechanicsExpectationEOM}) is now seen to be exactly in the form of a linear control problem, with the cavity photon number (or, equivalently, the laser power) playing the role of the control function with $u(t) = \langle \hat{a}^{\dagger}(t) \hat{a}(t) \rangle$.


\subsection{Circuit quantum electrodynamics and superconducting qubits}


Circuit quantum electrodynamics (QED) describes the interaction of linear and nonlinear superconducting circuits with quantized electromagnetic fields, usually in the microwave frequency range \cite{gu2017,blais2021}. This field has given rise to many new interesting phenomena in microwave photonics, as it is capable of reaching the ultra-strong coupling regime. Furthermore, when coupled with nonlinear elements such as Josephson junctions, consisting of an insulating break in a superconducting circuit, it allows the creation of a variety of types of superconducting qubits that are currently of intense interest as they are one of the most promising approaches that allowing development of scalable quantum computers \cite{nakamura1999}. 

The basic building block of these superconducting circuits is the LC circuit, which has the classical Hamiltonian
\begin{equation}
H_{LC} = \frac{Q^2}{2C} + \frac{\Phi}{2 L}
\end{equation}
where $L$ is the circuit inductance, $C$ is the capacitance, $Q$ is the charge on the capacitor, and $\Phi$ is the flux in the inductor. To quantize this system, one treats the charge and flux variables as non-commuting observables satisfying
\begin{equation}
[\hat{\Phi}, \hat{Q}] = i \hbar
\end{equation}
and creates bosonic operators $\hat{a}, \hat{a}^\dagger$ through the relations
\begin{equation}
\hat{\Phi} = \sqrt[4]{\frac{\hbar^2 L}{4 C}} \, (\hat{a} + \hat{a}^\dagger), \,\,\,\,\,\,\,\,\,
\hat{Q} = i \sqrt[4]{\frac{\hbar^2 C}{4 L}} \, (\hat{a}^{\dagger} - \hat{a}).
\end{equation} 
The resulting quantized Hamiltonian is given by \cite{blais2021}
\begin{equation}
\hat{H}_{LC} = \hbar \omega \left( \hat{a}^{\dagger} \hat{a} + \frac{1}{2} \right)
\label{eqHLC}
\end{equation}
with $\omega = 1/\sqrt{LC}$, and describes a single mode quantum harmonic oscillator. When the circuit is coupled to a classical voltage drive source at microwave frequencies, or equivalently a resonator consisting of a strong coherent microwave field with Rabi frequency $\Omega$, the Hamiltonian (\ref{eqHLC}) acquires an additional driving term of the form $\hat{H}_\text{d} = \Omega \hat{a} +\Omega^* \hat{a}^{\dagger}$. Finally, when arrays of these circuits are coupled via superconducting wiring or waveguides one obtains a set of linearly coupled harmonic oscillators described by the Hamiltonian
\begin{equation}
\frac{\hat{H}}{\hbar} = \sum_i \omega_i \hat{a}^{\dagger}_i \hat{a}_i + \sum_{i\neq j}\left( g_{ij} \hat{a}^{\dagger}_i \hat{a}_j +  g_{ij}^* \hat{a}^{\dagger}_j \hat{a}_i \right) + \sum_i \left( \Omega_i \hat{a} +\Omega_i \hat{a}^{\dagger} \right)
\label{eqCircuitQEDLinearH}
\end{equation}
where we have dropped the zero point energy terms. Eq.~(\ref{eqCircuitQEDLinearH}) is exactly the quadratic bosonic Hamiltonian we consider in this paper, without the squeezing terms. Consequently such a system is amenable to the control techniques we have described in this paper.

More generally, much of the interest in circuit QED arises due to the introduction of additional nonlinear circuit elements based on Josephson junctions. One of the most important of these circuits is the transmon, consisting of a Josephson junction with an internal capacitance. The Hamiltonian governing this system is given by \cite{gu2017,blais2021}
\begin{equation}
\hat{H}_T = 4 E_C \hat{n}^2 +\frac{1}{2} E_J \hat{\phi}^2 - E_J(\cos \hat{\phi} + \frac{1}{2} \hat{\phi}^2)
\label{eqFullTransmonH} 
\end{equation}
where $E_J$ is the (tunable) Josephson energy and $E_C$ is the charge energy inversely proportion to the capacitance. $\hat{n}$ is the charge number operator counting the number of charges on the capacitor, and $\hat{\phi}$ is the conjugate phase operator that is proportional to the flux operator describing the magnetic flux through the Josephson junction.

The transmon is usually analysed by introducing bosonic operators $\hat{b}$, $\hat{b}^{\dagger}$ given by
\begin{eqnarray}
\hat{\phi} &=& \left( \frac{2 E_C}{E_J} \right)^{1/4} (\hat{a} + \hat{a}^{\dagger}), \\
\hat{n} &=& \frac{i}{2} \left( \frac{E_J}{2 E_C} \right)^{1/4} (\hat{a}^{\dagger}_i - \hat{a}_i).
\end{eqnarray}
In the limit where $\hat{\phi}$ is ``small'', such that its expectation value remains near the minimum of the cosine term in (\ref{eqFullTransmonH}), $\hat{H}_T$ can be approximated by
\begin{equation}
H_T = \hbar \omega_q \hat{a}^{\dagger} \hat{a} - \frac{E_C}{2} \hat{a}^{\dagger} \hat{a}^{\dagger} \hat{a} \hat{a}
\end{equation}
where $\omega_q = \sqrt{8 E_C E_J}$. It is this nonlinearity that results in the energy levels of $\hat{H}_T$ no longer being evenly spaced, allowing the lowest two levels to be treated as an effective two-level system and used as a qubit. 

Just as in the quantized LC circuit case, transmons can be driven by an external voltage or microwave resonator which gives rise to a linear drive term, and coupled to other transmons or linear circuit elements. The form of the coupling depends on the regime the transmon is operated in and the bus architecture, but they can be described by the Hamiltonians of the form \cite{gu2017,blais2021,stehlik2021, hazra2021}
\begin{eqnarray}
\frac{H}{\hbar} &=&  \sum_i \left( \omega_i \hat{a}_i^{\dagger} \hat{a}_i - \frac{\delta}{2} \hat{a}_i^{\dagger} \hat{a}_i^{\dagger} \hat{a}_i \hat{a}_i \right) + \sum_{i\neq j} g_{ij} (\hat{a}^\dagger_i + \hat{a}_i)(\hat{a}^\dagger_j + \hat{a}_j) \nonumber \\
&&  + \sum_i \left( \Omega_i \hat{a}_i +\Omega_i \hat{a}^{\dagger}_i \right)
\label{eqQEDCoupledHamiltonian}
\end{eqnarray}
which, aside from the nonlinear quartic term, are also described by our formalism. This means that, in order for our control theory to be applicable, it is necessary that the nonlinear term is in some sense small. Specifically, if we wish to apply a set of controls that accomplish the state manipulation in total time $T$, we require $T \ll 1/\delta$. Fast control requires large control amplitudes, which is why it is an advantage that the control function amplitudes can be given by the magnitude of the external voltage drive. Voltage is relatively easy to apply precisely, can be changed rapidly, and can trivially have large amplitudes.

\begin{figure}[htp]
\includegraphics[width=1.0\columnwidth]{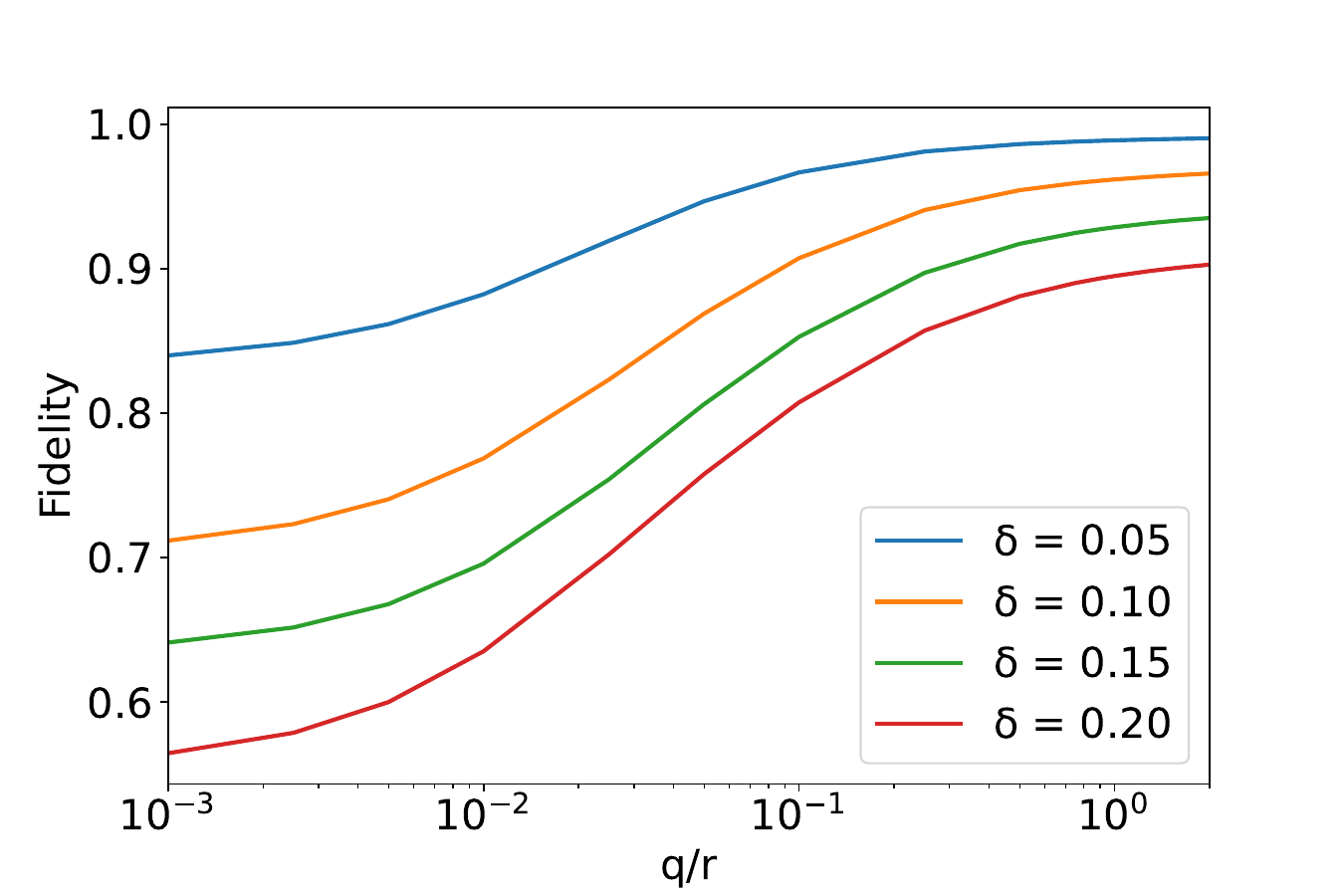}
\caption{Demonstration that optimal control can reduce fidelity errors arising from a strong nonlinearity $\delta$. We find a control function that moves a state of the (linear) one dimensional harmonic oscillator from $\langle x(0)\rangle = 0$, $\langle p(0)\rangle = 0$ to $\langle x(20)\rangle = 2.0$, $\langle p(20)\rangle = 0.5$ using optimal control to either prioritize staying close to the origin of the $(x,p)$ plane for as long as possible ($q/r$ large), or prioritizing keeping the control amplitudes low ($q/r$ small). In the absence of nonlinearities, the control would be perfect; we wish to see what happens if a nonlinearity is added. 
The plot shows the fidelity of the transport for the system described by the anharmonic oscillator Hamiltonian (\ref{eqAnharmonicOscillatorHamiltonianQoverR}), using the control function found for the linear harmonic oscillator case. We see that increasing the ratio $q/r$ results in higher fidelity results, due to the fact that the nonlinear error grows as $\langle x\rangle^2 + \langle p\rangle^2$ increases.}
\label{fig:figEffectOfNonlinearityOnFidelity}
\end{figure}

In addition, we can reduce the effect of the nonlinearity by applying the techniques of optimal control described in Section~\ref{secOptimalControl}. The effect of the nonlinear term increases as $\langle \hat{a} ^{\dagger} \hat{a}\rangle$ increases, or, equivalently, as the energy $\langle \hat{x}^2 \rangle + \langle \hat{p} ^2 \rangle$ increases. Consequently if one can find an optimal control that minimizes the energy for most of the control time, the effect of the nonlinearity will be suppressed. This can be accomplished by choosing a cost function
\begin{equation}
J = \int_0^T \left( q |{\mathbf{x}}(t)|^2 + r |u(t)|^2 \right) \, dt
\end{equation}
where $\mathbf{x} = (x,p)$ and changing the relative weights of $q$ and $p$. As the ratio $q/r$ increases, we are increasingly penalizing high energy excursions away from $x=p=0$ at the expense of allowing large control amplitudes, resulting in a control path that minimizes the integrated energy. As an example, we consider the single mode Hamiltonian
\begin{equation}
H = a^\dagger a + \frac{\delta}{2} a^\dagger a^\dagger a a + u(t)(a + a^\dagger).
\label{eqAnharmonicOscillatorHamiltonianQoverR}
\end{equation}
We first use our linear control scheme to find a control function $u(t)$ that maximizes the fidelity of nonadiabatically transporting a wave packet from $\langle x(0)\rangle = 0$, $\langle p(0)\rangle = 0$ to $\langle x(20)\rangle = 2.0$, $\langle p(20)\rangle = 0.5$ in the {\emph{absence}} of the nonlinearity, i.e. for $\delta = 0$, but using optimal control to ensure the control function minimizes the cost function. We then examine what happens to system evolution under that control function, but with a variety of nonlinearities applied. The result is shown in Figure~\ref{fig:figEffectOfNonlinearityOnFidelity} for a variety of nonlinear strengths, and demonstrates how the fidelity of wave function transport improves as we increase the ratio $q/r$. In effect, the optimal control framework allows us to maximize fidelity in the presence of nonlinearities, while still using linear control theory.

\section{Conclusion}

We have investigated the multimode bosonic quadratic Hamiltonian, where all terms are bilinear in bosonic creation and annihilation operators, with the addition of a set of time-dependent terms that are linear in the creation and annihilation operators. These additional time-dependent drivings serve as external controls to the system.

As this situation is a superset of the ubiquitous quantum harmonic oscillator, it describes a very broad range of physical systems. The time-dependent linear drive is also one of the easiest and most common ways to interact with such a system, as it physically corresponds to a displacement of the harmonic potential, although it may manifest in different ways, such as the Rabi frequency of a laser field interacting with an atom or the voltage drive to a superconducting qubit.

We have provided a general solution to this type of system in terms of the time dependent control functions, as well as an explicit form of the solution for the case where the linear driving term is independent of time. Of greater interest, however, is the situation where we are interested in the expectation values of moments of annihilation and creation operators. In this case we have shown how the equations of motion can be mapped to a classical linear control problem, which is an extremely well-studied and well-understood field. This allows us to provide conditions on when the system is expectation-value controllable, i.e. the linear drive allows complete control over any time-dependent quadrature expectation values, as well as an explicit recipe to construct control functions that accomplish this. When mapping this linear control theory back to the quantum problem, we see that the action of the linear coupling term in the Hamiltonian provides the freedom to provide arbitrary displacement operations on the bosonic modes.

Since determination of controllability depends on the exact form of the Hamiltonian, general statements about controllability must necessarily be broad. We provide a general result based on the eigenspectrum of the Hamiltonian, as well as the various strength of the linear mode couplings. Going beyond this requires
 a case-by-case analysis of a specific Hamiltonian and its symmetries. We provide an illustration of how this can work by considering a linear chain of quantum systems, with each coupled to its two nearest neighbours by both hopping and squeezing terms, and assume that our linear control term can only affect the system at one end of the chain. We show that for this system, for almost all sets of intra-chain couplings, it is possible to control all elements of the chain.
 
The ability to map the bilinear Hamiltonian with linear drive to classical linear control theory also opens up the possibility of considering optimal control. Optimal control investigates how a system can be controlled under some set of constraints, or by finding solutions that perform the desired control while simultaneously minimizing some cost function. We have shown how our formalism allows the mapping of the the bilinear Hamiltonian to classical optimal control problems, and illustrated its use on a chain of nearest-neighbour coupled quantum systems, demonstrating how position and momentum of the chain can be controlled while minimizing a parameterised cost function.

To demonstrate the utility of our framework, we have considered how it can be applied to a number of different systems. Specifically we have considered distortion-free transport of atomic wave packets beyond the adiabatic limit, the implementation of an echoed conditional displacement (ECD) gate, optomechanical systems, and circuit QED.

The example of wave packet transport considers the common case of moving a trapped system, such as a trapped atom or a Bose-Einstein condensate, from one physical location to another, with the final wave function identical to the intial wave function, by shifting the potential center of the trap. We provide a recipe for producing control functions that accomplish this transport in arbitrarily short times, while also respecting a cost function, and give a specific example.

The ECD gate example considers the recent interest in enhancing couplings between a cavity mode and a qubit, as this can allow QND readout of superconducting qubits or produce a gate that displaces the cavity mode conditional on the qubit state. We have shown how our formalism not only allows such a gate to be realised in short times with a choice of continuous (i.e. not delta-function or piecewise constant) control pulses, but also how arbitary conditional displacements of the cavity field can be achieved.

The optomechanical example is of interest as it shows how one can control the center of mass position and momentum of a micromechanical oscillator by modifying the intensity of an applied laser. One is not limited merely to the control of a single centre of mass mode, however. Provided there is some coupling between mechanical modes, which is almost always the case, our formalism shows when it is possible to manipulate all the modes individually via the single control of laser field intensity, and provides explicit solutions for the required time-dependence of that field.

As the last example, we considered the application of our framework to circuit QED, and specialize to the case of superconducing qubits based on transmons. We examined how the position and momentum states of a qubit can be shifted, as might be required, for example, when implementing GKP error correction. While our framework is linear, and such qubits include a nonlinearity, we show how one can use optimal control techniques to perform the operations while minimizing the effect of that nonlinearity, allowing our linear control solutions to perform well.

Finally, we stress that these examples are a merely a subset of what our framework can address, chosen to illustrate what can be accomplished. A bilinear multimode Hamiltonian with a linear driving term covers a large number of different types of quantum systems. The ability to understand when those systems are expectation-value controllable, being able to find those controls, and the option to minimize cost functions associated with those controls ensures our methods are broadly applicable.

\begin{acknowledgments}
We acknowledge helpful discussions with Thomas Volz regarding possible applications of this work. This research was funded in part by the Australian Research Council (project numbers FT190100106, DP210101367, CE170100009).
\end{acknowledgments}

\appendix

\section{Proof of the Kalman controllability criterion and explicit control solutions}
\label{secAppendixKalmanProof}

Suppose we have a linear system described by 
\begin{equation}
\frac{d}{dt} {\mathbf{x}}(t) = A {\mathbf{x}}(t)  + C {\mathbf{u}}(t)
\label{eqLinearControlDefinition1}
\end{equation}
where $A$ is an $n\times n$ matrix and $B$ is an $n\times m$ matrix, so that $A: \mathbb{C}^n \rightarrow \mathbb{C}^n$, \, $C: \mathbb{C}^m \rightarrow \mathbb{C}^n$.  ${\mathbf{x}(t)} \in \mathbb{C}^n$ and ${\mathbf{u}}(t) \in \mathbb{C}^m$ are vectors, where the ${\mathbf{x}}(t)$ denote the $n$ system configuration variables at time $t$, and the ${\mathbf{u}}(t)$ denote the $m$ continuous scalar control functions. As shown in Section~\ref{secQHOandNotation}, the unique solution to  Eq.~(\ref{eqLinearControlDefinition1}) is given by
\begin{equation}
{\mathbf{x}}(T) = e^{A T} {\mathbf{x}}(0) + \int_0^T e^{A (T-s) } C {\mathbf{u}}(s) \, ds.
\label{eqAppendixCauchySolution} 
\end{equation}

A system that evolves according to Eq.~(\ref{eqLinearControlDefinition1}) is called controllable at time $T>0$ if we can move any initial state ${\mathbf{x}}(0)$ to any other goal target state ${\mathbf{x}}(T) = {\mathbf{g}}$ at time $T$ by choice of suitable control functions ${\mathbf{u}}(t)$.

We wish to prove that the system (\ref{eqLinearControlDefinition1}) is controllable if and only if the Kalman matrix $K$ has rank $n$, where $K$ is constructed via
\begin{equation}
K =: \left[ B \,\,\, AC \,\, A^2C \,\,\, \dots A^{n-1}C  \,\, \right].
\label{eqAppendixKalmanDefinition}
\end{equation}
This matrix $K$ has size $n\times mn$.

Our proof is based on the one given in \cite{zabczyk2020}, although we extend the analysis from a purely real-valued matrix $A$ and control matrix $B$ to their complex-valued counterparts, as this is required for our quantum control problem. The proof consists of two parts. We show that if $K$ does not have rank $n$ the system is not controllable, and then we show that if $K$ does have rank $n$ we can find an explicit set of functions ${\mathbf{u}}(t)$ that perform the desired control.

First, we show that if $K$ does not have rank $n$ the system is not controllable at any time $T$. To do this, assume that the matrix $K$ has rank $r<n$, and perform a singular value decomposition of $K$. Since $r<n$ there must be at least one singular value equal to zero and an associated right singular vector ${\mathbf{v}} \in \mathbb{C}^n$ satisfying
\begin{equation}
K^{\dagger} {\mathbf{v}}  = {\mathbf{0}}
\label{eqRightSingularValue}
\end{equation}
where ${\mathbf{0}}$ denotes a zero vector. From the construction of $K$ in (\ref{eqAppendixKalmanDefinition}), and taking the Hermitian transpose of Eq.~(\ref{eqRightSingularValue}), we obtain
\begin{equation}
{\mathbf{v}}^{\dagger} \left[ C \,\,\, AC \,\, A^2C \,\,\, \dots A^{n-1}C  \,\, \right] = {\mathbf{0}}^T
\end{equation}
giving us the result that
\begin{equation}
{\mathbf{v}}^{\dagger} A^k C = {\mathbf{0}}^T \,\,\,\,\, \forall k\in \{ 0, 1, \dots n-1 \}.
\label{eqAppendixZeroColumns}
\end{equation}

The Cayley-Hamiltonian theorem tells us that any analytic function of a matrix $A$ of dimension $n$ can be written as a weighted sum of matrix powers of $A$ no higher than $n-1$. Specifically it shows that there exists a set of coefficients $c_k$ such that the matrix exponential is given by
\begin{equation}
\exp(A t) = \sum_{k=0}^{n-1} c_k(t) A^k.
\label{eqAppendixMatrixExp}
\end{equation}
From (\ref{eqAppendixZeroColumns}) and (\ref{eqAppendixMatrixExp}) we obtain
\begin{eqnarray}
{\mathbf{v}}^{\dagger}  \exp (A t) C &=& \sum_{k=0}^{n-1} c_k(t) {\mathbf{v}}^{\dagger}   A^k C \nonumber \\
&=& {\mathbf{0}}
\label{eqCHresult1}
\end{eqnarray}
for {\emph{any}} time $t$.

We choose our goal state ${\mathbf{g}}$ to the be the right singular vector ${\mathbf{v}}$ with singular value zero, and choose our starting vector to be the zero vector, i.e. ${\mathbf{x}}(0) = {\mathbf{0}}$. If we consider the overlap between our goal state and the system state ${\mathbf{x}}(t)$, then using the Cauchy solution (\ref{eqAppendixCauchySolution}) we obtain
\begin{eqnarray}
{\mathbf{g}}^T {\mathbf{x}}(T) &=&  {\mathbf{g}}^T  e^{A T} {\mathbf{x}}(0) + \int_0^T {\mathbf{g}}^T e^{A (T-s) } C {\mathbf{u}}(s) \, ds \nonumber \\
&=& 0
\label{eqAppendixOverlap1}
\end{eqnarray}
with the first term vanishing due to our choice of initial condition, and the second term vanishing due to Eq.~(\ref{eqCHresult1}). Eq.~(\ref{eqAppendixOverlap1}) shows that there is no there is no possible set of control functions ${\mathbf{u}}(t)$ that will take ${\mathbf{x}}(0)$ to ${\mathbf{g}}$, demonstrating that if $K$ does not have rank $n$, then the system is not controllable.

For the second part of the proof we show that if $K$ has rank $n$ we can find explicit functions ${\bf{u}}(t)$ that perform the desired control. Specifically, we wish to move ${\bf{x}}(0)$ to any other arbitrary goal target state ${\bf{x}}(T) = {\bf{g}}$ at time $T$ by choice of suitable control functions ${\bf{u}}(t)$.

We begin by noting that if $K$ has rank $n$, then there exists an $mn\times n$ matrix $\bar{K}$ with the property $K\bar{K} = \mathbb{1}_n$, where $\mathbb{1}_n$ denotes the $n\times n$ identity matrix. From the form of (\ref{eqAppendixKalmanDefinition}) this is equivalent to the statement that there exists a set of $n$, $m \times n$ matrices $\bar{K}_1,\ldots , \bar{K_n}$ satisfying
\begin{equation}
B \bar{K}_1 + A C \bar{K}_2 + A^2 C \bar{K}_3 + \cdots +  A^{n-1} C \bar{K}_n = \mathbb{1}_n.
\label{eqABKbarIdentity}
\end{equation}

Next we define a function $\mu(s)$ defined on the interval $[0,T]$ with the following properties:
\begin{itemize}
\item The first $n-1$ derivatives of $\mu(s)$ are continuous
\item $\frac{d^l}{ds^{l}} \mu(s) = 0$ at $s=0$ and $s=T$ for $l=0,1,\ldots , n-1$
\item $\int_0^T \mu(s) \,dt = 1$.
\end{itemize}

Finally, we introduce an auxillary vector of functions ${\mathbf{r}}(s)$ given by 
\begin{equation}
{\mathbf{r}}(s) = \mu(s) e^{A(s-T)} \left( {\mathbf{g}} -e^{AT}  {\mathbf{x}}(0) \right).
\end{equation}

With these definitions, we claim that the control functions
\begin{equation}
{\mathbf{u}}(t) = \sum_{l=1}^n \bar{K}_l  \frac{d^{l-1}}{dt^{l-1}} {\mathbf{r}}(t)
\end{equation} 
are the ones we require. 

We prove this by inserting the explicitly constructed ${\mathbf{u}}(t)$ into the Cauchy solution given by Eq.~(\ref{eqAppendixCauchySolution}). We obtain
\begin{equation}
{\mathbf{x}}(T) = e^{AT} {\mathbf{x}}(0) + \sum_{l=1}^n \int_0^T e^{A(T-s) } C  \bar{K}_l \frac{d^{l-1}}{ds^{l-1}} {\mathbf{r}}(s) \,ds.
\end{equation}
If we let 
\begin{eqnarray}
u &=& e^{A(T-s)} \\
dv &=&  \bar{K}_l \frac{d^{l-1}}{dt^{l-1}} {\mathbf{r}}(s) ds
\end{eqnarray}
and utilize $l-1$ applications of integration by parts we obtain
\begin{equation}
{\mathbf{x}}(T) = e^{AT} {\mathbf{x}}(0) + \sum_{l=1}^n \int_0^T e^{A(T-s) } A^{l-1} C  \bar{K}_l {\mathbf{r}}(s) \,ds.
\label{eqAppendixKalmanDerivation2}
\end{equation}
Applying Eq.~(\ref{eqABKbarIdentity}) this becomes
\begin{eqnarray}
{\mathbf{x}}(T) &=& e^{AT} {\mathbf{x}}(0) + \int_0^T e^{A(T-s) } {\mathbf{r}}(s) \,ds \nonumber \\
&=& e^{AT} {\mathbf{x}}(0) + \int_0^T e^{A(T-s) }  \mu(s) e^{A(s-T)} \left( {\mathbf{g}} -e^{AT} {\mathbf{x}}(0) \right) \, ds \nonumber \\
&=& {\mathbf{g}}
\end{eqnarray}
as required.

Finally, we note that in the case where there is only a single control function ${\mathbf{u}}(t) = u(t)$, which is the case for some examples considered in this paper, the Kalman matrix $K$ is square and $n \times n$. In this case the Kalman rank criterion (the system is controllable if $K$ has rank $n$) is equivalent to $K$ being invertible, which may in some situations be a simpler test.

\end{document}